\documentclass[amsmath,amssymb,12pt]{article}
\usepackage{jheppub}
\usepackage{hyperref}
\usepackage{graphicx}    
\newcommand{\vphi}[0]{\delta\phi}

\newcommand{\dlg}[0]{\lp g}

\newcommand{\virt}[0]{\hat{\delta}}
\newcommand{\Dvphi}[1]{\nabla_{#1}\lp\phi}
\newcommand{\DDvphi}[2]{\nabla_{#1}\nabla_{#2}\lp\phi}

\newcommand{\half}[0]{\frac{1}{2}}
\newcommand{\cs}[3]{\Gamma^{#1}_{\,\,\,\, #2#3}}

\newcommand{\ld}[0]{\mathcal{L}}

\newcommand{\dd}[0]{\textrm{d}}
\newcommand{\defn}[0]{\equiv}
\newcommand{\diag}[0]{\textrm{diag}}
\newcommand{\qsubrm}[2]{{#1}_{\scriptsize{\textrm{#2}}}}
\newcommand{\qsuprm}[2]{{#1}^{\textrm{#2}}}

\newcommand{\sol}[0]{\ld_{\scriptscriptstyle\{2\}}}

\newcommand{\tis}[0]{ {\theta}^{\scriptscriptstyle\rm{S}}}
\newcommand{\pis}[0]{ {\Pi}^{\scriptscriptstyle\rm{S}}}

\newcommand{\rbm}[1]{{\bf{#1}}}
\newcommand{\ci}[0]{\textrm{i}}
\newcommand{\kin}[0]{{\mathcal{X}}}
\newcommand{\hct}[0]{\mathcal{H}}
\renewcommand{\figurename}{Figure}
\newcommand{\ep}[0]{{ {\delta}_{\scriptscriptstyle{\rm{E}}}}}
\newcommand{\lp}[0]{{ {\delta}_{\scriptscriptstyle{\rm{L}}}}}
\def\be{\begin{equation}}
\def\ee{\end{equation}}
\def\bea{\begin{eqnarray}}
\def\eea{\end{eqnarray}}
\def\bse{\begin{subequations}}
\def\ese{\end{subequations}}
\newcommand{\lied}[1]{\pounds_{#1}}

\let\oldsqrt\sqrt
\def\sqrt{\mathpalette\DHLhksqrt}
\def\DHLhksqrt#1#2{%
\setbox0=\hbox{$#1\oldsqrt{#2\,}$}\dimen0=\ht0
\advance\dimen0-0.2\ht0
\setbox2=\hbox{\vrule height\ht0 depth -\dimen0}%
{\box0\lower0.4pt\box2}}

\newcommand{\sbm}[2]{#1_{\mathbb{#2}}}

\renewcommand{\tis}{\theta}
\subheader{\today}

\title{Computing model independent perturbations in dark energy and modified gravity}
\author[a]{Richard A. Battye}
\author[b]{and Jonathan A. Pearson}
\affiliation[a]{Jodrell Bank Centre for Astrophysics, School of Physics and Astronomy, The University of Manchester, Manchester M13 9PL, U.K}
\affiliation[b]{Department of Mathematical Sciences, Durham University, South Road, Durham, DH1 3LE, U.K}
\emailAdd{richard.battye@manchester.ac.uk}
\emailAdd{jonathan.pearson@durham.ac.uk}

\abstract{We present a methodology for computing model independent perturbations in dark energy and modified gravity. This is done from the Lagrangian for perturbations, by showing how   field content, symmetries, and physical principles are often sufficient ingredients for closing the set of perturbed fluid equations. The fluid equations   close once   ``equations of state for perturbations'' are identified: these are linear combinations of fluid and metric perturbations which construct gauge invariant entropy and anisotropic stress perturbations for broad classes of theories. Our main results are the proof of the equation of state for perturbations presented in a previous paper, and the development of the required calculational tools. 
 }
 
\begin{document}
\maketitle

\section{Introduction}
Since the discovery of apparent cosmic acceleration \cite{Perlmutter:1998np, Riess:1998cb, Riess:1998dv, Frieman:2008sn} there has been an explosion in the number  of dark energy \cite{Copeland:2006wr, dakrenergy_amendola} and modified gravity theories \cite{Clifton:2011jh}   constructed in an attempt to describe these observations.  The   route model builders  usually go down is to write a Lagrangian at background order according to some phenomenological or physically motivated principles, obtain constraints at background order on the theory, perturb it and obtain further constraints from the perturbations. This entire process is   model dependent, with the results and constraints obtained being limited to the theoretical prejudices which were imposed by the functional form of the Lagrangian which was written down.  The proliferation of models  has prompted  recent interest in looking for ways to phenomenologically parameterize theories \cite{0004-637X-506-2-485, Weller:2003hw, PhysRevD.69.083503, PhysRevD.76.104043,   PhysRevD.77.103524,  Skordis:2008vt,  PhysRevD.81.083534, PhysRevD.81.104023,  Appleby:2010dx, Hojjati:2011ix,  Thomas:2011sf, Zuntz:2011aq, Baker:2011jy, Baker:2011wt, Bloomfield:2011np, Bloomfield:2012ff,  Mueller:2012kb, Kunz:2012aw, Baker:2012zs, Bloomfield:2013efa, Baker:2013hia, 2013arXiv1310.6026F}.   Constructing a good set of phenomenological tools and probes of perturbations in the dark sector is particularly pertinent given the  recent data releases from {\textit{CFHTLenS}} \cite{Heymans:2012gg}, \textit{Planck} \cite{Planck:2013kta, Ade:2013tyw} and in the future, the Dark Energy Survey \cite{Abbott:2005bi}, \textit{LSST} \cite{Abate:2012za}, and  \textit{Euclid} \cite{Amendola:2012ys}.

The formalism we introduced in \cite{Battye:2012eu, Pearson:2012kb, BattyePearson_connections, PearsonBattye:eos, Bloomfield:2013cyf}, and develop in the current paper,  does not require a Lagrangian for the theory to be presented for useful and consistent information about the dark sector to be extracted from observations. Our formalism can be thought of as a way to phenomenologically parameterize deviations of the gravity theory realized by nature from General Relativity. This can be done with   specific theories in mind, or by studying the signatures of generic theories. The important point is that we  obtain consistent cosmological perturbations from a model independent formalism: we are able to remain agnostic about the functional form of the Lagrangian.

The way in which     the problem is tackled is caught between a tension of ``\textit{theoretical generality}'' and ``\textit{experimental feasibility}''. From a theorists perspective, generality is key; however, this usually results in a system with more freedom than it is reasonable to expect observations to be able to constrain. Our strategy is therefore to study general theories which are imposed with (often well motivated) restrictions, whilst retaining important features of the general theory. 

%
%

The key aspect to our approach is how  we ``package'' the parameterization. The new ``PPF'' approach, outlined in \cite{Skordis:2008vt, Baker:2011jy, Baker:2012zs},  provides the general modifications to the gravitational field equations. The free functions in the modifications are called the PPF functions. There are a large number of ``free'' PPF functions for a general theory, but particular theories may   severely restrict the form and freedom of the coefficients. In spirit,  our approach is similar since we  identify all the PPF functions for    modified gravity theories satisfying various restrictions.  Our additional contribution to this is to  provide a useful way to package the modifications,  by characterizing \textit{equations of state for dark sector perturbations}.

Our aim in this paper  is to extend the formalism we introduced in \cite{Battye:2012eu, Pearson:2012kb, BattyePearson_connections, PearsonBattye:eos} for parameterizing dark sector perturbations to encompass  substantially broader classes of theories (see also  \cite{PhysRevD.76.023005}). This paper also acts as companion to \cite{PearsonBattye:eos}: here we explain, justify, and prove the claims made in that short paper. Our particular aims can be summarized as
\begin{itemize}
\item Present general modifications to gravitational field equations that are relevant for ``high derivative'' scalar field theories, in a model independent way.
\item Understand how to impose reparameterization invariance.
\item Obtain an understanding of how different field contents of theories affect observables, via equations of state for dark sector perturbations.
\item Motivate these modifications from an action for perturbations. This action for perturbations can be calculated from an explicit theory.
\end{itemize}

The idea is to modify the Einstein-Hilbert action with a term which contains all non-standard gravitational physics; we call this term the \textit{dark Lagrangian}. This modified action is written as
\bea
S = \int \dd^4x\, \sqrt{-g} \, \bigg[ \frac{R}{16\pi G} - \qsubrm{\ld}{matter} -  \qsubrm{\ld}{d}\bigg].
\eea
Varying the action with respect to the metric $g_{\mu\nu}$ gives 
\bea
G_{\mu\nu} = 8 \pi G \big[ T_{\mu\nu} + U_{\mu\nu}\big].
\eea
All contributions due to the dark Lagrangian $\qsubrm{\ld}{d}$ are contained within the \textit{dark energy-momentum tensor} $U_{\mu\nu}$. We assume that the energy-momentum tensor that comes from the matter Lagrangian is conserved, $\nabla_{\mu}{T^{\mu\nu}}=0$, which immediately implies that the dark energy-momentum tensor is also conserved
\bea
\nabla_{\mu}{U^{\mu\nu}}=0.
\eea
The field equations for perturbations are
\bea
\ep G_{\mu\nu} = 8 \pi G \big[ \ep T_{\mu\nu} + \ep U_{\mu\nu}\big],
\eea
where ``$\ep$'' is the relevant perturbation operator (we will explain why it has the ``E'' subscript later on). The perturbed conservation equation is
\bea
\ep(\nabla_{\mu}U^{\mu\nu})=0.
\eea
The goal of this paper is to elucidate how different field contents of the dark Lagrangian can influence the gravitational field equations at perturbed order, whilst assuming an absolute minimum of theoretical structure for the Lagrangian of the dark sector; this will tell us how to construct the perturbed dark energy-momentum tensor $ \ep U_{\mu\nu}$. We are able to obtain a ``usefully small'' number of free functions which can     be constrained with current observational data.

\textit{\textbf{Setup of the background and notation}} We will assume that the geometry of the background space-time is spatially homogeneous and isotropic,     this is described by a spatially flat FRW metric.  This is written in conformal coordinates as $g_{\mu\nu} = a^2(\tau)\eta_{\mu\nu}$, where $\eta_{\mu\nu} = \diag(-1,1,1,1)$ is the Minkowski metric. The symmetry of the background enables us to use a $(3+1)$ decomposition: we foliate the space with 3D hypersurfaces whose metric is $\gamma_{\mu\nu}$. The 3D surfaces are peirced by a time-like unit vector $u_{\mu}$. The metric is thus decomposed as $g_{\mu\nu} = \gamma_{\mu\nu} - u_{\mu}u_{\nu}$, where $u_{\mu}$ and  $\gamma_{\mu\nu}$   are subject to the conditions that
\bea
u^{\mu}u_{\mu} = -1,\qquad u^{\mu}\gamma_{\mu\nu}=0,\qquad \gamma_{\mu\nu} = \gamma_{(\mu\nu)}.
\eea
An orthogonal vector $V_{\mu}$ is a vector that satisfies $u^{\mu}V_{\mu}=0$. We will make use of the transverse-traceless orthogonal projection operator,
\bea
\label{eq:sec:defn-perp}
{\perp^{ \alpha\beta}}_{ \mu\nu} \defn {\gamma^{\alpha}}_{\mu}{\gamma^{\beta}}_{\nu} - \tfrac{1}{3}\gamma^{\alpha\beta}\gamma_{\mu\nu}.
\eea
This operator satisfies
\bea
u^{\mu}{\perp^{ \alpha\beta}}_{ \mu\nu} =0,\qquad \gamma^{\mu\nu}{\perp^{ \alpha\beta}}_{ \mu\nu} =0,\qquad {\perp^{ \alpha\beta}}_{ \mu\nu} {\perp^{\mu\nu}}_{\rho\sigma}= {\perp^{\alpha\beta}}_{\rho\sigma}.
\eea
The space-time covariant derivative of $u_{\mu}$ defines the extrinsic curvature tensor $K_{\mu\nu}$ of the 3D sheets,
\bea
K_{\mu\nu} = {\gamma^{\alpha}}_{\mu}{\gamma^{\beta}}_{\nu}K_{\alpha\beta} = K_{(\mu\nu)} = \nabla_{\mu}u_{\nu} = \tfrac{1}{3}K\gamma_{\mu\nu},\qquad K \defn {K^{\mu}}_{\mu} = \gamma^{\mu\nu} K_{\mu\nu}.
\eea
We use an overdot to denote derivative along $u_{\mu}$, and an overline above the derivative to denote spatial differentiation. That is, for some quantity  $X_{\nu}$,
\bea
\dot{X}_{\nu} \defn u^{\mu}\nabla_{\mu}X_{\nu},\qquad \bar{\nabla}_{\mu}X_{\nu} \defn {\gamma^{\alpha}}_{\mu}\nabla_{\alpha}X_{\nu}.
\eea
\section{Fluid language}
Rather than follow the usual route and cast the parameterization in terms of ``fields'', we use the more physically intuitive ``fluid'' description. This is a useful way to collect  all modifications to each component of the gravitational field equations. For instance, only certain derivatives and combinations of fields in the underlying dark sector theory will go into modifying the sources of   given components  of the perturbed gravitational field equations.

This approach is already commonly used at the level of the cosmological background. The dark energy momentum tensor  $U_{\mu\nu}$ has just  two components: the density, $\rho$, and pressure, $P$, of the dark fluid. These {\textit{macro}scopic fluid quantities contain the observationally relevant parts of the \textit{micro}scopic dark sector Lagrangian (if the background spacetime is FRW). The dark energy-momentum tensor is simply written as
\bea
U_{\mu\nu} = \rho u_{\mu}u_{\nu} + P\gamma_{\mu\nu}
\eea
and  satisfies the conservation equation $\nabla_{\mu}{U^{\mu}}_{\nu}=0$, whose only component is  $\dot{\rho} = - 3\hct(\rho+P)$.
The system of background field equations is  not yet closed, unless the pressure $P$ is specified in terms of field variables which have evolution equations. The most common way to do this is to write the equation of state $P = w\rho$, where in general $w = w(a)$. With this equation of state the background field equations close.  This is the only piece of freedom at the background which a dark sector theory can  modify.

At the level of linearized perturbations, the components of the (Eulerian) perturbed dark energy-momentum tensor can be parameterized as
\bea
\label{eq:sec:ep_u_flui-intro}
\ep {U^{\mu}}_{\nu} = \delta\rho u^{\mu}u_{\nu} + 2 (\rho+P)v^{(\mu}u_{\nu)} + \delta P {\gamma^{\mu}}_{\nu} + P{\Pi^{\mu}}_{\nu}.
\eea
The perturbation operator ``$\ep$'' will be explained in the next section, but for now it should just be understood to be the relevant perturbation for the perturbed gravitational field equations.
The components $\delta\rho, v^{\mu}, \delta P$ and ${\Pi^{\mu}}_{\nu}$ are the dark sector perturbed density, velocity, perturbed pressure and anisotropic stress: these are the \textit{perturbed fluid variables} of the dark sector.
Explicitly, each of the perturbed fluid variables can be found from a given expression for $\ep {U^{\mu}}_{\nu}$  by applying projectors along various directions,
\bse
\label{eq:sec:pert-fld-vars}
\bea
\delta\rho &=& u_{\mu}u^{\nu}\ep {U^{\mu}}_{\nu},\\
(\rho+P)v^{\alpha} &=& -u_{\mu}{\gamma^{\alpha}}_{\nu}\ep {U^{\mu}}_{\nu},\\
\delta P &=& \tfrac{1}{3}{\gamma_{\mu}}^{\nu} \ep {U^{\mu}}_{\nu},\\
P\Pi^{\alpha\beta} &=& {\perp^{\alpha\beta}}_{\mu}{}^{\nu}\ep {U^{\mu}}_{\nu}.
\eea  
\ese
Most commonly, $\ep {U^{\mu}}_{\nu}$  will be computed or given in terms of perturbed field variables (such as metric  or scalar field perturbations); (\ref{eq:sec:pert-fld-vars}) can   be used to determine how these field variables combine to construct the fluid variables -- we will give explicit examples later on.

The components of (\ref{eq:sec:ep_u_flui-intro}) are constrained by the perturbed conservation equation
\bea
\label{eq:sec:pett-cons-eq-el-preexpandout}
\ep(\nabla_{\mu}{U^{\mu}}_{\nu})=0.
\eea
This has two independent projections, which, using (\ref{eq:sec:ep_u_flui-intro}),   respectively become
\bse
\label{eq:sec:pert-fld-eqns-gengag}
\bea
\dot{\delta\rho} + K(\delta\rho + \delta P) + (\rho+P)\bar{\nabla}_{\mu}v^{\mu} + \rho u^{\alpha}\ep \cs{\mu}{\alpha}{\mu} + u^{\nu}{U^{\mu}}_{\alpha}\ep\cs{\alpha}{\mu}{\nu}=0,
\eea
\bea
&&(\rho+P)  \dot{v}_{\alpha}   + [\dot{\rho} + \dot{P} + \tfrac{4}{3}K(\rho+P)]v_{\alpha} + \bar{\nabla}_{\alpha}\delta P + P{\gamma^{\beta}}_{\alpha}\bar{\nabla}_{\lambda}{\Pi^{\lambda}}_{\beta} \nonumber\\
&&\qquad\qquad+ P{\gamma^{\mu}}_{\alpha}\ep \cs{\beta}{\beta}{\mu} - {\gamma^{\lambda}}_{\alpha}{U^{\mu}}_{\beta}\ep \cs{\beta}{\mu}{\lambda}=0,
\eea
\ese
where the perturbation to the Christoffel symbols is given by
\bea
\label{eq:pert-christsymb}
\ep\cs{\alpha}{\mu}{\nu} = \tfrac{1}{2}g^{\alpha\beta} \big( \nabla_{\mu}\ep g_{\nu\beta} + \nabla_{\nu}\ep g_{\mu\beta} - \nabla_{\beta}\ep g_{\mu\nu}\big).
\eea
What we see, therefore, is that the   perturbed conservation equation (\ref{eq:sec:pett-cons-eq-el-preexpandout}) provides evolution equations for two of the perturbed fluid variables: the density perturbation $    \delta\rho$ and the velocity field $v^{\alpha}$  (the perturbed metric variables which will come out from the perturbed Christoffel symbols (\ref{eq:pert-christsymb}) are evolved via the gravitational field equations). However, the set of perturbed fluid equations (\ref{eq:sec:pert-fld-eqns-gengag}) are not closed since there is no evolution equation for the perturbed pressure $\delta P$ or the anisotropic stress ${\Pi^{\mu}}_{\nu}$. This is highlighted much more clearly in the synchronous gauge and Fourier space and for scalar perturbations only,   since (\ref{eq:sec:pert-fld-eqns-gengag}) becomes
\bse
\label{eq:sec:fluid-eqs-fourier}
\bea
\left( \frac{\delta}{1+w}\right)^{\cdot} = - \big[ - k^2\theta + \tfrac{1}{2}\dot{h}\big] - \frac{3 \hct}{1+w} w\Gamma,
\eea
\bea
(1+w) {\dot{\theta}}{ } &=& - \hct (1+w)\left(1-3\frac{\dd P}{\dd\rho}\right)\theta -\frac{\dd P}{\dd\rho}\delta-  {w}{} \Gamma+\tfrac{2}{3}w\Pi ,
\eea
\ese
where the gauge invariant entropy perturbation 
\bea
w\Gamma\defn  \bigg(\frac{\delta P}{\delta\rho} - \frac{\dd P}{\dd\rho} \bigg)\delta 
\eea
 is used to  package the pressure perturbation. We have defined the scalar velocity field, $\theta$, via $\theta = \ci \rbm{k}\cdot\rbm{v}/k^2$. The scalar metric perturbations, $h$ (and below we will use $\eta$) are defined as in \cite{Ma:1994dv}. Notice that this fluid is general, in the sense that we have allowed for non-zero entropy perturbations, anisotropic stress, and $\dot{w}\neq 0$.

It should now be clear that all that needs to be specified is the entropy perturbation $w\Gamma$ and the anisotropic stress ${\Pi^{\mu}}_{\nu}$ of the dark fluid: these are the two ``physical'' pieces of freedom which a dark sector theory will end up specifying. Once these are provided in terms of   variables whose equations of motion are already specified, the system of equations closes and can be solved. These will be key in the packaging of our parameterization, and will form what we call the \textit{equations of state for perturbations}. Schematically, these equations of state for perturbations look like
\bea
w\Gamma = A_1\delta + A_2\theta + A_3 \dot{h} + \ldots,\qquad \Pi = B_1\delta + B_2\theta + B_3 \eta + \ldots,
\eea
where $\{A_i, B_i\}$ represent the free functions which control the precise form of the equations of state for perturbations. If the underlying theory is reparameterization invariant, these functions must form a gauge invariant combination (since $w\Gamma$ and $\Pi$ are both gauge invariant by definition).

The key point which will come out of our analysis is that $w\Gamma$ and ${\Pi^{\mu}}_{\nu}$ are constructed from dynamical fluid and metric components in different ways depending on the   field content and symmetries of the dark sector theory. The most pertinent question our approach is able to answer is precisely which of these dynamical components  are required  to construct the gauge invariant entropy perturbation and the anisotropic stress  to describe broad classes of modified gravity and dark energy theories. 

\section{Perturbed EMT from field content}
We will now describe how   knowing the field content of the dark sector   is sufficient for obtaining the perturbed dark energy-momentum tensor from the Lagrangian for perturbations. We then discuss issues of reparameterization invariance and provide field equations.

\subsection{The Lagrangian for perturbations}
\label{append:conn-lag}
We will start off with a very general theory, where the field content of the dark sector includes the metric $g_{\mu\nu}$ and a scalar field $\phi$, as well as the  \textit{partial} derivatives of these fields. The dark sector field content that  we study is 
\bea
\label{eq:sec:append-sosft0lag-field-content}
\ld = \ld(g_{\mu\nu}, \partial_{\alpha}g_{\mu\nu}, \phi, \partial_{\alpha}\phi , \partial_{\alpha}\partial_{\beta}\phi).
\eea
Note that we have not included the second partial derivative of the metric: it is clear how to extend the framework presented here to include such field contents. The Lagrangian for perturbations in this theory is given by everything quadratic in the first perturbation to these field variables, yielding
\bea
\label{eq:sol-2osft}
\sol &=& \mathcal{A}\lp\phi^2 + \mathcal{B}^{\mu}\lp\phi \Dvphi{\mu} + \tfrac{1}{2} \mathcal{C}^{\mu\nu}\Dvphi{\mu}\Dvphi{\nu} + \mathcal{D}^{\mu\nu}\lp\phi\DDvphi{\mu}{\nu} \nonumber\\
&&  + \mathcal{E}^{\mu\alpha\beta}\Dvphi{\mu}\DDvphi{\alpha}{\beta} + \tfrac{1}{2} \mathcal{F}^{\mu\nu\alpha\beta}\DDvphi{\mu}{\nu}\DDvphi{\alpha}{\beta}\nonumber\\
&&+{\mathcal{I}_{}}^{\rho\mu\nu} \nabla_{\rho}\dlg_{\mu\nu} \lp\phi + {\mathcal{J}_{}}^{\rho\mu\nu\alpha}  {\nabla}_{\rho}\dlg_{\mu\nu}\Dvphi{\alpha} \nonumber\\
&&+ {\mathcal{N}_{}}^{\rho\mu\nu\alpha\beta} {\nabla}_{\rho}\dlg_{\mu\nu}\DDvphi{\alpha}{\beta} + \tfrac{1}{2} {{{\mathcal{M}_{}}^{\rho\mu\nu}}_{}}^{\sigma\alpha\beta}{\nabla}_{\rho}\dlg_{\mu\nu}\nabla_{\sigma}\dlg_{\alpha\beta}\nonumber\\
&&+ \tfrac{1}{4}\big[ \mathcal{V}^{\mu\nu}\lp\phi \dlg_{\mu\nu} + \mathcal{Y}^{\alpha\mu\nu}\dlg_{\mu\nu}\Dvphi{\alpha} + \mathcal{Z}^{\mu\nu\alpha\beta}\dlg_{\alpha\beta} \DDvphi{\mu}{\nu}\nonumber\\
&&\qquad + \tfrac{1}{2} \mathcal{W}^{\mu\nu\alpha\beta}\dlg_{\mu\nu}\dlg_{\alpha\beta} + {\mathcal{U}_{}}^{\rho\mu\nu\alpha\beta} \nabla_{\rho}\dlg_{\mu\nu}\dlg_{\alpha\beta}\big] .
\eea
The perturbation operator ``$\lp$'' in (\ref{eq:sol-2osft}) will be explained shortly, but for now it should simply be taken as a perturbation operator. 
There are 15  tensors $\{\mathcal{A}, \dots, \mathcal{Z}^{\mu\nu\alpha\beta}\}$ in the Lagrangian for perturbations, each describing   couplings between perturbed field variables. For this reason, we call the tensors \textit{coupling tensors}. The coupling tensors are  functions of background field variables  only; in the cosmological background, this means that the coupling tensors are   functions of time  and not position. In addition, they have a number of symmetries which can be deduced from the objects that they are contracted with. For example, since $\lp g_{\mu\nu} = \lp g_{(\mu\nu)}$ and $\DDvphi{\mu}{\nu}  = \DDvphi{(\mu}{\nu)}$, one can deduce that 
\bea
\mathcal{D}^{\mu\nu} = \mathcal{D}^{(\mu\nu)},\qquad \mathcal{W}^{\mu\nu\alpha\beta} = \mathcal{W}^{(\mu\nu)(\alpha\beta)} = \mathcal{W}^{\alpha\beta\mu\nu},\qquad \mathcal{Z}^{\mu\nu\alpha\beta} = \mathcal{Z}^{(\mu\nu)(\alpha\beta)}.
\eea
This is not an exhaustive list, and   symmetries of the other coupling tensors can be read off from (\ref{eq:sol-2osft}).

Providing the Lagrangian for perturbations is sufficient for calculating the linearized field equations,
\bea
\lp G_{\mu\nu} = 8 \pi G \lp T_{\mu\nu} + \lp U_{\mu\nu},
\eea
where the perturbed dark energy momentum tensor $ \lp U_{\mu\nu}$ is calculated from $\sol$ via
\bea
\label{eq:sec:lp_u_howtocompute}
\lp U^{\mu\nu} = - \half \bigg[4\frac{\virt}{\virt \lp g_{\mu\nu}}\sol + U^{\mu\nu} g^{\alpha\beta} \lp g_{\alpha\beta}\bigg].
\eea
Here, ``$\virt$'' denotes functional variation.
Clearly, $\sol$ contains more information than is needed for the linearized gravitational field equations.  The perturbed dark energy momentum tensor of all theories with field content (\ref{eq:sec:append-sosft0lag-field-content}) can be constructed from (\ref{eq:sol-2osft}) by using (\ref{eq:sec:lp_u_howtocompute}), and subsequntly written as
\bse
\label{eq:sec:sosft_lp_u-op}
\bea
\label{eq:sec:sosft_lp_u-op-a}
\lp U^{\mu\nu} = \hat{\mathbb{Y}}^{\mu\nu}\lp\phi + \hat{\mathbb{W}}^{\mu\nu\alpha\beta} \lp g_{\alpha\beta},
\eea
where $\hat{\mathbb{Y}}^{\mu\nu}$ and  $\hat{\mathbb{W}}^{\mu\nu\alpha\beta}$ are derivative operators that are given by
\bea
\hat{\mathbb{Y}}^{\mu\nu}&\defn& \mathbb{A}^{\mu\nu}  + \mathbb{B}^{\alpha\mu\nu}\nabla_{\alpha}  + \mathbb{C}^{\alpha\beta\mu\nu} \nabla_{\alpha}\nabla_{\beta}  + \mathbb{D}^{\rho\alpha\beta\mu\nu}\nabla_{\rho}\nabla_{\alpha}\nabla_{\beta} ,\\
\hat{\mathbb{W}}^{\mu\nu\alpha\beta} &\defn& \mathbb{E}^{\mu\nu\alpha\beta}  + \mathbb{F}^{\rho\mu\nu\alpha\beta}\nabla_{\rho}+\mathbb{G}^{\rho\sigma\mu\nu\alpha\beta}\nabla_{\rho}\nabla_{\sigma}  ,
\eea
\ese
where we have defined
\bse
\label{eq:sec:defs-sosft-bold-bbm}
\bea
\mathbb{A}^{\mu\nu}  &\defn& - \tfrac{1}{2} \big[ \mathcal{V}^{\mu\nu}   -4\nabla_{\rho}\mathcal{I}^{\rho\mu\nu}\big],\\
\mathbb{B}^{\alpha\mu\nu}  &\defn& - \tfrac{1}{2} \big[\mathcal{Y}^{\alpha\mu\nu}   -4 (\mathcal{I}^{\alpha\mu\nu}+\nabla_{\rho}\mathcal{J}^{\rho\mu\nu\alpha})\big],\\
\mathbb{C}^{\alpha\beta\mu\nu}  &\defn& - \tfrac{1}{2} \big[\mathcal{Z}^{\alpha\beta\mu\nu} -4 (\mathcal{J}^{\beta\mu\nu\alpha}+\nabla_{\rho}\mathcal{N}^{\rho\mu\nu\alpha\beta}) \big],\\
\mathbb{D}^{\rho\alpha\beta\mu\nu}  &\defn& 2\mathcal{N}^{\rho\mu\nu\alpha\beta}  ,\\
\mathbb{E}^{\mu\nu\alpha\beta}   &\defn& - \tfrac{1}{2} \big[\mathcal{W}^{\mu\nu\alpha\beta} + U^{\mu\nu}g^{\alpha\beta} -\nabla_{\rho}\mathcal{U}^{\rho\mu\nu\alpha\beta} \big],\\
\label{eq:sec:f-defn-sol}
\mathbb{F}^{\rho\mu\nu\alpha\beta}  &\defn& - \tfrac{1}{2} \big[ {\mathcal{U}_{}}^{\rho\alpha\beta\mu\nu}  -\mathcal{U}^{\rho\mu\nu\alpha\beta}  -4 \nabla_{\epsilon}\mathcal{M}^{\epsilon\mu\nu\rho\alpha\beta}\big],\\
\mathbb{G}^{\rho\sigma\mu\nu\alpha\beta}  &\defn& 2 \mathcal{M}^{\rho\mu\nu\sigma\alpha\beta}  .
\eea
\ese
The expressions (\ref{eq:sec:defs-sosft-bold-bbm}) provides us with an understanding as to how the coupling tensors in the Lagrangian for perturbations combine to construct the perturbed energy-momentum tensor; these relationships will prove to be crucial when it comes to understanding the structure of its components.

In the subsequent analysis we will restrict ourselves to a subset of these theories: only those which are linear in $\partial_{\alpha}g_{\mu\nu}$. This has the consequence of removing all quadratic couplings of the derivative of the perturbed metric in the Lagrangian for perturbations. That is, it sets $ \mathcal{M} =0$ in $\sol$ and therefore $\mathbb{G} =0$ in $\lp U^{\mu\nu}$. There is no reason in principle to prevent the inclusion of such tensors, but this restriction significantly simplifies the algebra. Notice that a corollary of this is that from (\ref{eq:sec:f-defn-sol}) we see that  $\mathbb{F}^{\rho\mu\nu\alpha\beta}  = - \mathbb{F}^{\rho\alpha\beta\mu\nu}$. An anti-symmetry of this type could not have been realized without having the underlying structure of the Lagrangian for perturbations from which the perturbed energy-momentum tensor was derived.

We call the bold-face tensors $\{\mathbb{A}, \ldots, \mathbb{G}\}$ used in (\ref{eq:sec:sosft_lp_u-op}) the \textit{EMT expansion tensors}. The indices in the EMT expansion tensors  in $\hat{\mathbb{W}}$ are structured so that the last two are contracted with $\lp g_{\alpha\beta}$ (and so  are  symmetric), the next two are the same indices on $\lp U^{\mu\nu}$ (and   are still symmetric),   and the first indices are contracted with covariant derivatives (and have no symmetry). In general,  the EMT expansion tensors have the following symmetries in their indices:
\bse
\label{eq:sec:tensors-field-gen-sosft}
\bea
\mathbb{A}^{\mu\nu} = \mathbb{A}^{(\mu\nu)} ,\quad \mathbb{B}^{\alpha\mu\nu} = \mathbb{B}^{\alpha(\mu\nu)},\quad \mathbb{C}^{\alpha\beta\mu\nu} =  \mathbb{C}^{(\alpha\beta)(\mu\nu)},\quad\mathbb{D}^{\rho\alpha\beta\mu\nu} =  \mathbb{D}^{\rho(\alpha\beta)(\mu\nu)},
\eea
\bea
\mathbb{E}^{\alpha\beta\mu\nu} =  \mathbb{E}^{(\alpha\beta)(\mu\nu)},\quad\mathbb{F}^{\rho\alpha\beta\mu\nu} =  \mathbb{F}^{\rho(\alpha\beta)(\mu\nu)},\quad\mathbb{G}^{\rho\sigma\alpha\beta\mu\nu} =  \mathbb{G}^{\rho\sigma(\alpha\beta)(\mu\nu)}.
\eea
\ese
Note that $\mathbb{E}$ has the same symmetries as $\mathbb{C}$, and $\mathbb{F}$ has the same symmetries as $\mathbb{D}$. In backgrounds with ``arbitrary'' symmetry   these tensors have a very large number of free components; later on we will impose   the background to be spatially isotropic, which substantially reduces the number of independent components of these tensors.

To show that a given explicit theory (e.g. one written down from a background Lagrangian) fits into a particular flavour of our formalism, it suffices to show that its Lagrangian for perturbations is of the form (\ref{eq:sol-2osft}), and that is guaranteed if its field content is given by (\ref{eq:sec:append-sosft0lag-field-content}). The   theory (\ref{eq:sol-2osft}) will contain Lorentz violating theories and theories which do not satisfy reparameterization invariance.   One of our aims is to identify the maximum possible freedom in theories of the type (\ref{eq:sol-2osft}). We will then identify the freedom for reasonable subsets of theories, since retaining too much generality yields a highly intractable set of equations; we are constantly keeping in mind the desire to use observationally obtained data to constrain the space of allowed theories. This will yield expressions from which we can extract the ``dark sources'' to the linearized gravitational field equations.

\subsection{Reparameterization invariance}
As it stands, the perturbed dark energy-momentum tensor (\ref{eq:sec:sosft_lp_u-op}) will be able to describe  very wide classes of theories,   including those which are usually deemed to be theoretically unattractive. One of the  properties we might like a theory to possess is an invariance under reparameterization,
\bea
\label{reparap-defn}
x^{\mu} \rightarrow x^{\mu} + \xi^{\mu}.
\eea
Linearized gravitational theories, of the types considered in this paper, are not \textit{a priori} reparameterization invariant (RI). For example, under (\ref{reparap-defn}) the metric perturbation $\delta g_{\mu\nu}$  transforms as
\bea
\delta g_{\mu\nu} \rightarrow \delta g_{\mu\nu} + 2\nabla_{(\mu}\xi_{\nu)}.
\eea
One common tactic is to build gauge invariant cosmological perturbation theory by constructing the theory from gauge invariant perturbed field variables.

We are able to \textit{impose} reparameterization invariance on the theory, which corresponds to imposing  constraints and relationships between the components of the EMT expansion tensors (\ref{eq:sec:tensors-field-gen-sosft}). To do this we need to understand the role that the reparameterization-field $\xi^{\mu}$   plays in the system. This is done by writing all expressions in their ``reparameterized'' form which involves relating  the perturbation operators $\lp$ and $\ep$. These correspond to perturbations in   Lagrangian and Eulerian coordinate systems respectively. For a   field variable $X$ say, these perturbations are linked via 
\bea
\lp X = \ep X + \lied{\xi}{X},
\eea
 where $\lied{\xi}$ is the Lie derivative along $\xi^{\mu}$ (the vector which generates coordinate reparameterizations).  For the current purposes it is useful to think of $\xi^{\mu}$ as being a \textit{Stuckelberg field}, whose role is to restore reparameterization invariance, and therefore to think of $\lp g_{\mu\nu}$ as being the Stuckelberg-completed (and thus RI) metric perturbation. For  the scalar field and metric    perturbations, and the perturbed dark energy-momentum tensor one  has
\bse
\bea
\lp \phi &=&\ep \phi +\lied{\xi}\phi, \\
\lp g_{\mu\nu}&=&\ep g_{\mu\nu} +\lied{\xi}g_{\mu\nu}, \\
\lp U^{\mu\nu}&=&\ep U^{\mu\nu}  +\lied{\xi}U^{\mu\nu}, 
\eea
where the Lie derivatives are
\bea
\lied{\xi}\phi &=&  \xi^{\mu}\nabla_{\mu}\phi,\\
\lied{\xi}g_{\mu\nu} &=&  2 \nabla_{(\mu}\xi_{\nu)},\\
\lied{\xi}U^{\mu\nu} &=&  \xi^{\alpha}\nabla_{\alpha}U^{\mu\nu} - 2 U^{\alpha(\mu}\nabla_{\alpha}\xi^{\nu)}.
\eea
\ese
Putting these expressions  together, and using (\ref{eq:sec:sosft_lp_u-op-a}) to provide the Lagrangian perturbed dark energy-momentum tensor, the Eulerian perturbed dark  energy-momentum tensor which sources the gravitational field equation is
\bea
\label{eq:sec:ep-emt-ywd}
\ep U^{\mu\nu} &=& \hat{\mathbb{Y}}^{\mu\nu}\ep\phi + \hat{\mathbb{W}}^{\mu\nu\alpha\beta} \ep g_{\alpha\beta}+\Delta_{\{\xi\}}\ep U^{\mu\nu},
\eea
where the contribution due to the Stuckelberg field $\xi^{\mu}$ is
\bea
\label{eq:sec:cont-stru-general}
\Delta_{\{\xi\}}\ep U^{\mu\nu}\defn\hat{\mathbb{Y}}^{\mu\nu}\lied{\xi}\phi + \hat{\mathbb{W}}^{\mu\nu\alpha\beta} \lied{\xi} g_{\alpha\beta}- \lied{\xi}U^{\mu\nu}.
\eea
Applying the projectors (\ref{eq:sec:pert-fld-vars}) onto (\ref{eq:sec:ep-emt-ywd})   provides expressions for the perturbed fluid variables in terms of perturbed field variables. Explicitly, one   obtains
\bse
\bea
\delta\rho&=&u_{\mu}u_{\nu}\ep U^{\mu\nu} -\rho u^{\mu}  u^{\nu}\ep g_{\mu\nu},\\
(\rho+P)v^{\alpha} &=&-u_{\mu}{\gamma^{\alpha}}_{\nu}\ep U^{\mu\nu} +\rho {\gamma^{\alpha\mu}} u^{\nu}\ep g_{\mu\nu},\\
\delta P &=&\tfrac{1}{3} {\gamma}_{\mu\nu}  \ep U^{\mu\nu} +  \tfrac{1}{3}P\gamma^{\mu\nu}\ep g_{\mu\nu},\\
P\Pi^{ \alpha\beta} &=& {\perp^{  \alpha\beta}}_{\mu\nu} \ep U^{\mu\nu} +  P{\perp^{  \alpha\beta\nu\lambda}}  \ep g_{\lambda\nu}.
\eea
\ese
The extra terms on the right-hand-side are   due to the fact that   the variation operator does not commute with index raising and lowering. 

\textit{A priori} all components of the   Stuckelberg field $\xi^{\mu}$ are dynamical and couple to the perturbed gravitational field equations via $\Delta_{\{\xi\}}\ep U^{\mu\nu}$. We will shortly provide their equations of motion. Only when $\Delta_{\{\xi\}}\ep U^{\mu\nu}$ is independent of a given component of $\xi^{\mu}$ is the theory invariant under reparameterizations of that relevant space-time coordinate. That is, if $\xi^0$ does not appear   in any components of $\Delta_{\{\xi\}}\ep U^{\mu\nu}$ then the theory is $SO(1,0)$ invariant (i.e. under time reparameterizations), and if $\xi^i$ does not appear then the theory is $SO(0,3)$ invariant (i.e. under spatial reparameterizations). Finally, if neither $\xi^0$ nor $\xi^i$ appear in $\Delta_{\{\xi\}}\ep U^{\mu\nu}$, then the theory is fully $SO(1,3)$ reparameterization invariant. Later on we will show precisely how the components of the EMT expansion tensors can be arranged to make each of these invariances manifest.

\subsection{Perturbed conservation equation}

Providing the perturbed dark energy-momentum tensor  is only part of the story. We also require that $\ep U^{\mu\nu}$ satisfies a conservation equation,
\bea
\label{eq:Sec:eu-cons-eq}
\ep (\nabla_{\mu}U^{\mu\nu})=0.
\eea
Using (\ref{eq:sec:ep-emt-ywd}) for $\ep {U^{\mu\nu}}$, this can be written  schematically to show the  contributions to (\ref{eq:Sec:eu-cons-eq}) from the perturbed scalar field, $F^{\nu}$,  from the perturbed metric, $  J^{\nu}$, and   from the $\xi$-field, $E^{\nu}$,
\bea
F^{\nu} =   J^{\nu} +   E^{\nu},
\eea
where
\bse
\bea
F^{\nu}&\defn& \nabla_{\mu}(\hat{\mathbb{Y}}^{\mu\nu}\ep\phi),
\\
  J^{\nu} &\defn& - \big[\nabla_{\mu}(\hat{\mathbb{W}}^{\mu\nu\alpha\beta} \ep g_{\alpha\beta}) + 2 U^{\alpha(\mu}\ep \cs{\nu)}{\mu}{\alpha}\big],
\\
E^{\nu} &\defn&   \nabla_{\mu}(\lied{\xi}U^{\mu\nu}) - \nabla_{\mu}(\hat{\mathbb{Y}}^{\mu\nu}\lied{\xi}\phi) - \nabla_{\mu}(\hat{\mathbb{W}}^{\mu\nu\alpha\beta}\lied{\xi}g_{\alpha\beta}).
\eea
\ese
We now see that (\ref{eq:Sec:eu-cons-eq}) constitutes the equation of motion of the Stuckelberg fields.
It should be clear that constraints must be placed on the components of $\hat{\mathbb{Y}}$ and $\hat{\mathbb{W}}$ (and therefore on the EMT expansion tensors) to keep these equations of motion at most of second order.

\section{The perturbed fluid variables}
\label{sec:schem_summ_perfvars}
In this section   we provide the perturbed fluid variables as functions of the perturbed field variables for a ``generic'' theory. This will tell us exactly how time and space derivatives of \textit{field} variables combine to construct the \textit{fluid} variables; remembering that it is actually the fluid variables which source the gravitational field equations governing the evolution of the perturbed metric variables.

In the appendix we provide detailed descriptions of the calculations  performed to obtain the perturbed fluid variables for a subset of the theories described by (\ref{eq:sec:sosft_lp_u-op}). The subset is the set of theories which  
\begin{itemize}
\item [(a)]have second order field equations,
\item [(b)] are at most linear in $\partial_{\alpha}g_{\mu\nu}$, and
\item [(c)]are reparameterization invariant.
\end{itemize}
Whilst     condition (a) is  not  likely to be relaxed, conditions (b) and (c) can be relaxed, but we won't explicitly do so in this paper (for the sake of ``simplicity'').  Condition (b)   means that  the perturbed dark energy-momentum tensor  is given by (\ref{eq:sec:sosft_lp_u-op}) where the  $\hat{\mathbb{W}}$ tensor is expanded to
\bea
\hat{\mathbb{W}}^{\mu\nu\alpha\beta} &=& \mathbb{E}^{\mu\nu\alpha\beta}  + \mathbb{F}^{\rho\mu\nu\alpha\beta}\nabla_{\rho} .
\eea
Demanding reparameterization invariance translates  into the requirement that the gauge fields contribution to the perturbed fluid variables vanishes, that is,
\bse
\bea
&u_{\mu}u_{\nu} \Delta_{\{\xi\}} \ep U^{\mu\nu} = 0,\qquad u_{\mu}{\gamma^{\alpha}}_{\nu}\Delta_{\{\xi\}} \ep U^{\mu\nu}=0,&\\
& \gamma_{\mu\nu}\Delta_{\{\xi\}} \ep U^{\mu\nu}=0,\qquad {\perp^{\alpha\beta}}_{\mu\nu}\Delta_{\{\xi\}} \ep U^{\mu\nu}=0,&
\eea
\ese
where $ \Delta_{\{\xi\}} \ep U^{\mu\nu} $ is given by (\ref{eq:sec:cont-stru-general}).

To resolve these conditions to such an extent that the perturbed fluid variables can be written down as known functions of the perturbed field variables  requires very dense and involved calculations and is presented in Appendix \ref{append:calc-pfvs}.  The calculation is formulated entirely in tensorial notation, and so one can   obtain a clear and unambiguous  understanding of the geometrical meaning of reparameterization invariance and precisely how to impose second order field equations. 

The result of the calculation  is that the perturbed fluid variables for the \textit{subset} of the theories described by (\ref{eq:sec:sosft_lp_u-op}) which satisfy conditions (a-c) above, are given by
\bea
\label{eq:sec:fluid-vsrs-ppf}
\left( \begin{array}{c} \delta-A_{14}\dot{h} \\ \tis \\ \delta P  \end{array}\right) = \left( \begin{array}{ccc} \qsubrm{A}{11} & \qsubrm{A}{12} & 0 \\ \qsubrm{A}{21} & \qsubrm{A}{22} & 0  \\ \qsubrm{A}{31} & \qsubrm{A}{32} & \qsubrm{A}{33} \end{array}\right)\left( \begin{array}{c} \vphi \\ \dot{\vphi} \\ \ddot{\vphi} \end{array}\right) ,
\eea
 and all have zero scalar anisotropic stress, $\pis=0$ (in addition, the vector and tensor anisotropic stresses   vanish). One finds that  all $\qsubrm{A}{IJ}$ are scale independent (that is, they just depend on time and not scale $k$). The matrix $[\qsubrm{A}{IJ}]$ is called the \textit{activation matrix}. We reiterate that we have not specified the functional form of the background Lagrangian: only its field content and various symmetry requirements.
 
There are other classes of theories which have non-vanishing $\pis,  {\Pi}^{\scriptscriptstyle\rm{V}}$ and $ {\Pi}^{\scriptscriptstyle\rm{T}}$ that are constructed in this model independent way, notably the elastic dark energy theory \cite{PhysRevD.76.023005}.

\section{Equations of state for dark sector perturbations}
At the level of the cosmological background, despite their complexity,  all dark theories  boil down to specifying the  time dependence of a single function, which is commonly thought of as the equation of state parameter, $w(a)$.  Clearly, different theories predict different values and functional forms of $w(a)$,  but that is all they do: there is nothing else to be measured at the background that will tell us about the nature of the dark sector.  An obvious question then arises: how many functions need to be measured to characterize perturbations in the dark sector?

In \cite{Battye:2012eu} we showed that the   cosmological perturbations of    all reparameterization invariant single derivative scalar field theories (i.e. scalar field theories of the type $\ld = \ld(\phi, \kin)$, where $\kin \defn - \half g^{\mu\nu}\nabla_{\mu}\phi\nabla_{\nu}\phi$ is the kinetic scalar) are   encoded by a single   function, which we called $\alpha$     (this function   is, in general, time-dependent). This function arose as a single parameter in an \textit{equation of state for dark sector perturbations} (similar ``closure relations'' have also since been given in \cite{Sawicki:2012re}). In analogue to $w(a)$ at   background order, wide varieties of theories may well give rise to the same values of $\alpha$, in which case these theories will be indistinguishable at the level of linearized perturbations. The point is that observationally all we can hope to do is constrain the values of $\alpha$   (at the level of linearized perturbations). A series of questions naturally arise. For instance: what do the equations of state for dark sector perturbations look like for more general theories? Which fluid and metric variables appear in the equations of state? Specifically, those theories containing more than one derivative of the scalar field and/or derivatives of the metric.

The contributions to the fluid variables (\ref{eq:sec:fluid-vsrs-ppf}) from $\vphi, \dot{\vphi}$ and $\ddot{\vphi}$  in $\delta P$ will introduce terms which   \textit{a priori} require another equation of motion and are thus not-closed.  To remove these non-closed terms we   derive equations of state. We will now show how to compute the equation of state for perturbations from the activation matrix (\ref{eq:sec:fluid-vsrs-ppf}).

We start off by writing down the following part of the activation matrix which contains the known fluid variables:
\bea
\label{eq:sec:subexp-actmat}
\binom{\delta - A_{14}\dot{h}}{\theta}= \left( \begin{array}{cc}A_{11} & A_{12} \\ A_{21} & A_{22}\end{array}\right) \binom{\vphi}{\dot{\vphi}}  .
\eea
We obtain   expressions for $\vphi, \dot{\vphi}$ and $\ddot{\vphi}$ by inverting and differentiating (\ref{eq:sec:subexp-actmat}) and isolating the combination $\dot{\delta} - 3 \hct(1+w)\dot{\theta}$. This process yields
\bse
\label{eq:sec:vphi-dp-ddp-solved}
\bea
\vphi &=& \frac{1}{\mathcal{D}}\big[A_{22}(\delta - A_{14}\dot{h}) - A_{12}\theta\big],\\
\dot{\vphi} &=& \frac{1}{\mathcal{D}} \big[ A_{11}\theta - A_{21}(\delta - A_{14}\dot{h})\big],\\
\ddot{\vphi} &=& \frac{1}{\mathcal{E}} \big[ \dot{\delta} - 3 \hct(1+w)\dot{\theta} - \dot{A}_{14}\dot{h} - A_{14}\ddot{h} - \mathcal{F}\vphi - \mathcal{G}\dot{\vphi}\big],
\eea
\ese
where we defined the denominators as
\bse
\bea
\mathcal{D} &\defn& A_{11}A_{22} - A_{12}A_{21},\\
\mathcal{E} &\defn& A_{12} - 3 \hct(1+w) A_{22},
\eea
\ese
and the numerators as
\bse
\bea
\mathcal{F} &\defn& \dot{A}_{11} - 3 \hct(1+w) \dot{A}_{21},\\
\mathcal{G} &\defn& A_{11} + \dot{A}_{12} - 3\hct(1+w)(A_{21} + \dot{A}_{22}).
\eea
\ese
We now insert (\ref{eq:sec:vphi-dp-ddp-solved}) into $\delta P$'s row of the activation matrix (\ref{eq:sec:fluid-vsrs-ppf}) to obtain the following schematic form of the pressure perturbation:
\bea
\label{eq:sec:eltap=schem}
\delta P = \mathcal{A}_1 \delta + \mathcal{A}_2\theta + \mathcal{A}_3\dot{h} + \mathcal{A}_4 \ddot{h} + \mathcal{A}_5\big[\dot{\delta} - 3 \hct(1+w)\dot{\theta}\big].
\eea
The $\mathcal{A}_i$ are  defined in terms of the $\qsubrm{A}{IJ}$ as
\bse
\label{eq:sec:defn-as-script}
\bea
\mathcal{A}_1 &\defn& \frac{1}{\mathcal{D}} \bigg[ A_{22}\bigg( A_{31} - \frac{\mathcal{F}}{\mathcal{E}}A_{33}\bigg) - A_{21}\bigg(A_{32} - \frac{\mathcal{G}}{\mathcal{E}}A_{33}\bigg) \bigg],\\
\mathcal{A}_2 &\defn& \frac{1}{\mathcal{D}} \bigg[ A_{11}\bigg( A_{32} - \frac{\mathcal{G}}{\mathcal{E}}A_{33}\bigg) - A_{12} \bigg( A_{31} - \frac{\mathcal{F}}{\mathcal{E}}A_{33}\bigg) \bigg],\\
\mathcal{A}_3 &\defn& \frac{1}{\mathcal{D}} \bigg[ A_{21}A_{14}\bigg( A_{32} - \frac{\mathcal{G}}{\mathcal{E}}A_{33}\bigg) - A_{14}A_{22}\bigg( A_{31} - \frac{\mathcal{F}}{\mathcal{E}}A_{33}\bigg) - \frac{\mathcal{D}}{\mathcal{E}} A_{33}\dot{A}_{14} \bigg],\\
\mathcal{A}_4 &\defn& - \frac{1}{\mathcal{E}} A_{33}A_{14},\\
\mathcal{A}_5 &\defn& \frac{1}{\mathcal{E}} A_{33}.
\eea
\ese
We then use the perturbed fluid equations (\ref{eq:sec:fluid-eqs-fourier}) to replace the ``$\dot{\delta} - 3 \hct(1+w)\dot{\theta}$'' combination in  (\ref{eq:sec:eltap=schem}).
After doing this, one obtains the following schematic form of the entropy perturbation
\bea
\label{eq:sec:entrop-pregag-inv}
w\Gamma = \mathcal{B}_1 \delta + \mathcal{B}_2\theta + \mathcal{B}_3\dot{h} + \mathcal{B}_4\ddot{h},
\eea
where the $\mathcal{B}_{ i}$ are given by
\bse
\label{eq:sec:defn-b-script}
\bea
\rho\mathcal{B}_1 &\defn& \mathcal{A}_1 + 3 \hct w\mathcal{A}_5 - \tfrac{\dd P}{\dd\rho}\rho,\\
\rho\mathcal{B}_2 &\defn& \mathcal{A}_2 + (1+w)\big[3\hct^2(1-3\tfrac{\dd P}{\dd\rho}) + k^2\big]\mathcal{A}_5,\\
\rho\mathcal{B}_3 &\defn& \mathcal{A}_3 - \tfrac{1}{2} (1+w)\mathcal{A}_5,\\
\rho\mathcal{B}_4 &\defn& \mathcal{A}_4.
\eea
\ese
We now see that the only $\mathcal{B}_{ i}$ with scale dependence is $\mathcal{B}_2$, and that can be written as $\mathcal{B}_2  = \mathcal{B}_2^{(1)}(t) + \mathcal{B}_2^{(2)}(t)k^2$.
The entropy perturbation (\ref{eq:sec:entrop-pregag-inv}) now needs to take on gauge invariant form. In order to impose this, we recall that the fluid and metric variables transform from the synchronous to the conformal Newtonian gauge, defined as $\dd s^2 = a^2(\tau)\big[ -(1+2\Psi)\dd\tau^2 + (1-2\Phi)\dd\rbm{x}^2 \big]$, via
\bse
\label{eq:Sec:transf-syn-cn}
\bea
\delta &=& \hat{\delta} + 3 \hct(1+w)\zeta,\\
\theta &=& \hat{\theta} + \zeta,\\
\eta &=& \Phi + \hct\zeta,\\
\dot{h} &=& - 6 (\dot{\Phi} + \hct\Psi) +  \big[2k^2 - 6 (\dot{\hct} - \hct^2)\big]\zeta.
\eea
\ese
Here, $\zeta$ is the gauge transformation parameter and all gauge independent quantities need to be independent of $\zeta$. Additional transformations can be computed, making use of $\dot{\zeta} = \Psi - \hct\zeta$. 

We have a function, $w\Gamma$, constructed in the synchronous gauge   in (\ref{eq:sec:entrop-pregag-inv}), which we wish to put into gauge invariant form.  To do this, we use  (\ref{eq:Sec:transf-syn-cn}), to write $w\Gamma$ in the conformal Newtonian gauge:
\bea
w\Gamma &=& \mathcal{B}_1\hat{\delta} + \mathcal{B}_2\hat{\theta} -6\mathcal{B}_3(\dot{\Phi} + \hct\Psi) + \mathcal{B}_4\big[ - 6(\ddot{\Phi} + \hct\dot{\Psi}) + 2 k^2 - 12 \dot{\hct} + 6 \hct^2\big]\nonumber\\
&& \qquad\qquad+ \zeta\big[ 3\hct(1+w)\mathcal{B}_1 + \mathcal{B}_2+ 2 \mathcal{B}_3(k^2 - 3[\dot{\hct} - \hct^2] ) \nonumber\\
&&\qquad\qquad\qquad\qquad+ \mathcal{B}_4(-6\ddot{\hct} + 18 \hct\dot{\hct} - 6 \hct^3 - 2 k^2\hct) \big].
\eea
The last term in brackets multiplying $\zeta$ is required to vanish for $w\Gamma$ to be gauge invariant. We will pick particular forms of $\mathcal{B}_i$ which will satisfy this requirement and will yield a useful form of $w\Gamma$. 
From the outset we will define
\bse
\label{eq;sec:leaduptogaginventropu-flshf}
\bea
\mathcal{B}_1 \defn \alpha -\tfrac{\dd P}{\dd\rho}.
\eea
Suppose we had $\mathcal{B}_3 = \mathcal{B}_4=0$, then the choice $\mathcal{B}_2 = - 3\hct(1+w)\mathcal{B}_1$ would yield a gauge invariant function $w\Gamma$. This motivates us to define for the general case $\mathcal{B}_3\neq \mathcal{B}_4\neq 0$,
\bea
\mathcal{B}_2 \defn - 3\hct(1+w)\mathcal{B}_1\beta_1.
\eea
Similarly, from working out the required value of $\mathcal{B}_3$ in the case $\mathcal{B}_4=0$, we are motivated to define
\bea
\mathcal{B}_3 \defn - \frac{3\hct(1+w)\mathcal{B}_1\beta_2}{2k^2-6(\dot{\hct} - \hct^2)}.
\eea
In the full case where all terms are present, the only value of $\mathcal{B}_4$ which yields a gauge invariant combination is
\bea
\mathcal{B}_4 = \frac{3\hct(1+w)\mathcal{B}_1(1-\beta_1-\beta_2)}{6\ddot{\hct} + 6 \hct^3 - 18 \hct\dot{\hct} + 2 k^2\hct}.
\eea
\ese
Using (\ref{eq;sec:leaduptogaginventropu-flshf}),
 the entropy perturbation (\ref{eq:sec:entrop-pregag-inv}) becomes
\bea
\label{eq:sec:gag-inv-entr}
w\Gamma &=& (\alpha - \tfrac{\dd P}{\dd\rho})\bigg[ \delta - 3 \hct(1+w)\beta_1\theta - \frac{3\hct(1+w)\beta_2}{2k^2-6(\dot{\hct} - \hct^2)}\dot{h} \nonumber\\
&&\qquad\qquad\qquad\qquad\qquad+ \frac{3\hct(1+w)(1-\beta_1-\beta_2)}{6\ddot{\hct} + 6 \hct^3 - 18 \hct\dot{\hct} + 2 k^2\hct}\ddot{h}\bigg]. 
\eea
In the conformal Newtonian gauge, (\ref{eq:sec:gag-inv-entr}) becomes
\bea
w\Gamma &=& (\alpha - \tfrac{\dd P}{\dd\rho})\bigg[ \hat{\delta} - 3 \hct(1+w)\beta_1\hat{\theta} + \frac{9\hct(1+w)\beta_2}{k^2 - 3(\dot{\hct} - \hct^2)}(\dot{\Phi} + \hct\Psi) \nonumber\\
&&\qquad  - 3\hct(1+w)\frac{ 3 (\ddot{\Phi} + \hct\dot{\Psi})+(6\dot{\hct} - 3 \hct^2 - k^2)\Psi }{3\ddot{\hct} + 3 \hct^3- 9 \hct\dot{\hct}  + k^2\hct}(1-\beta_1-\beta_2)\bigg].
\eea
Equation (\ref{eq:sec:gag-inv-entr})  is the gauge-invariant entropy perturbation which closes the perturbed fluid equations (written in the synchronous gauge). There are three free dimensionless functions: $  \{\alpha, \beta_1, \beta_2\}$.  In a future paper \cite{BattyePearsonMoss_edeconstraints} we will confront the parameters in the equations of state with observational data.

One should note that the combinations (\ref{eq;sec:leaduptogaginventropu-flshf}) end up imposing
\bea
&&3\hct(1+w)\mathcal{A}_1 + \mathcal{A}_2 + \big( 2k^2 - 6[\dot{\hct} - \hct^2]\big) \mathcal{A}_3- \big(6\ddot{\hct} + 6 \hct^3 - 18\hct \dot{\hct} + 2 k^2\hct\big) \mathcal{A}_4\nonumber\\
&& \qquad\qquad\qquad\qquad  + 3(1+w)\big( \dot{\hct} + 3 \hct^2[w - \tfrac{\dd P}{\dd\rho}]\big)\mathcal{A}_5 = 3 \hct(1+w)\tfrac{\dd P}{\dd\rho}\rho 
\eea
on the $\mathcal{A}_i$ (\ref{eq:sec:defn-as-script}), and
\bea
\label{eq:sec:realtino-bi}
\mathcal{B}_4 = \frac{3\hct(1+w)\mathcal{B}_1 + \mathcal{B}_2 + \big( 2k^2 - 6[\dot{\hct} - \hct^2]\big)\mathcal{B}_3}{6\ddot{\hct} + 6 \hct^3 - 18\hct \dot{\hct} + 2 k^2\hct}
\eea
on the $\mathcal{B}_i$ (\ref{eq:sec:defn-b-script}). This corresponds to non-trivial relationships between the $\qsubrm{A}{IJ}$ (\ref{eq:sec:fluid-vsrs-ppf}). In the simple case where $A_{14} = A_{33}=0$, the condition (\ref{eq:sec:realtino-bi}) becomes $\mathcal{B}_2 = -3\hct(1+w)\mathcal{B}_1$, which can be verified to hold precisely for $k$-essence theories.

The important thing we have done is to compute the equations of state for perturbations without specifying the functional form of the dark sector Lagrangian. The equation of state (\ref{eq:sec:gag-inv-entr}) truly is model independent. It does depend, however, on the assumptions (a)-(c) outlined at the beginning of section \ref{sec:schem_summ_perfvars}.  

We will conclude this section with a short   example which does not satisfy reparameterization invariance. In previous work \cite{PhysRevD.76.023005, BattyePearson_connections} we studied the elastic dark energy theory, or equivalently a time-dependent massive gravity theory. In that theory, the dark sector Lagrangian is composed of the metric only,  and spatial reparameterization invariance is explicitly broken since they correspond to the deformations of an elastic medium. The equations of state for perturbations are given by $w\Gamma =0$ and
\bea
w\pis =\tfrac{3}{2}(w - \qsubrm{c}{s}^2) \times \left\{ \begin{array}{cc} \big[ \delta - 3(1+w)\eta\big] & \mbox{synchronous gauge},\\
 \big[ \delta - 3(1+w)\Phi\big] & \mbox{conformal Newtonian gauge}.\end{array}\right.
\eea
The (gauge invariant) combination ``$\delta -3 (1+w)\eta$'' arose naturally from the theory, even though spatial reparameterization invariance is explicitly broken, and $\qsubrm{c}{s}^2$ is the sound speed of the elastic medium. More general theories could lead to the inclusion of higher time-derivatives of $\eta$.

\section{Examples}
The results we presented in the previous sections were for ``general'' Lagrangians, where we only imposed the field content and reparameterization invariance and we never proposed a functional form of the Lagrangian. This yields expressions which hold for a very broad range of theories -- this could be percieved as a weakness.  What we can do, however, is to start from a more familiar standpoint, and write down the functional form of the Lagrangian. 

In this section we show that there is a relatively quick and easy way to compute the equation of state for perturbations for a theory with a specified Lagrangian, and indeed these are included within the general case (\ref{eq:sec:gag-inv-entr}).
\subsection{Minimally coupled scalar field theories}
As the first and simplest example, we will take the dark sector Lagrangian to be that for minimally coupled scalar fields:
\bea
\ld = \ld(\phi, \kin),
\eea
where $\kin \defn - \tfrac{1}{2}\nabla^{\mu}\phi\nabla_{\mu}\phi$. The energy density and pressure are given by 
\bea
\rho = 2\ld_{,\kin}\kin - \ld,\qquad  P=\ld,
\eea
which are functions with the following dependancies: 
\bea
\rho = \rho(\phi, \kin),  \qquad P = P(\phi, \kin).
\eea
The first variations of these functions is then given by
\bea
\delta\rho = \rho_{,\phi}\vphi + \rho_{,\kin}\delta\kin,\qquad \delta P = P_{,\phi}\vphi + P_{,\kin} \delta\kin.
\eea
For this theory it is simple to obtain $\theta = b_1\vphi$ (where $b_1 \defn -(2\kin)^{-1/2}$) and $\pis=0$. The activation matrix is thus
\bea
\left( \begin{array}{c} \delta\rho \\ \theta \\ \delta P \end{array}\right) = \left( \begin{array}{cc} \rho_{,\phi} & \rho_{,\kin} \\ b_1 & 0 \\ P_{,\phi} & P_{,\kin} \end{array}\right) \binom{\vphi}{\delta\kin}.
\eea
The perturbed field variables $\vphi, \delta\kin$ can be eliminated in favour of the perturbed fluid variables $\delta\rho, \theta$ to give
\bea
\vphi = \frac{1}{b_1}\theta,\qquad \delta\kin = \frac{1}{\rho_{,\kin}}\delta\rho - \frac{\rho_{,\phi}}{b_1\rho_{,\kin}}\theta.
\eea
The perturbed pressure can then be written in terms of ``known'' perturbed fluid variables,
\bea
\delta P =\frac{P_{,\kin}}{\rho_{,\kin}}\delta\rho + \frac{\rho_{,\phi}}{b_1}\bigg[ \frac{P_{,\phi}}{\rho_{\phi}} - \frac{P_{\kin}}{\rho_{,\kin}}\bigg]\theta
\eea
It is simple to show that the gauge invariant entropy perturbation is
\bea
\label{eq:sec:kess-case}
w\Gamma = (\alpha-\tfrac{\dd P}{\dd\rho})\big[\delta - 3 \hct(1+w)\theta\big],
\eea
with
\bea
\alpha \defn\frac{P_{,\kin}}{\rho_{,\kin}} =  \bigg( 1 + \frac{2\kin\ld_{,\kin\kin}}{\ld_{,\kin}}\bigg)^{-1}.
\eea
This has provided us with a well known result: the perturbed fluid equations for minimally coupled dark energy models close with a single parameter, $\alpha$. 
\subsection{Kinetic gravity braiding}
\label{sec:kgbeos}
The second example we consider  forms the first three terms of Horndeski's theory \cite{Horndeski:1974wa, Deffayet:2011gz, Kobayashi:2011nu}, and is called the Kinetic Gravity Braiding (KGB) theory \cite{Creminelli:2006xe, Creminelli:2008wc, Deffayet:2010qz, Pujolas:2011he}. This theory represents a useful   example of theories which end up introducing perturbed metric variables into the equation of state. The KGB Lagrangian is
\bea
\ld = \mathcal{A}(\phi, \kin)\square\phi + \mathcal{B}(\phi, \kin).
\eea
where $\kin \defn - \half g^{\mu\nu}\nabla_{\mu}\phi\nabla_{\nu}\phi$ as usual, and $\square\phi \defn \nabla^{\mu}\nabla_{\mu}\phi$. The energy-momentum tensor (EMT) is given by
\bea
\label{eq:sec:kgb-emt}
U_{\mu\nu} = \ld_{,\kin}\nabla_{\mu}\phi\nabla_{\nu}\phi  + 2 \nabla_{(\mu}\mathcal{A}\nabla_{\nu)}\phi + Pg_{\mu\nu},\qquad P \defn \mathcal{B} - \nabla^{\mu}\phi \nabla_{\mu}\mathcal{A}.
\eea
From  (\ref{eq:sec:kgb-emt}), the density $\rho$ and pressure $P$ for a spatially isotropic and homogeneous background are given by
\bse
\label{kgb_rhop}
\bea
\rho &=& -\mathcal{B} + 2(\mathcal{A}_{,\phi} + \mathcal{B}_{,\kin} )\kin - 2 \mathcal{A}_{,\kin}\kin\sqrt{2\kin}K,\\
P &=& \mathcal{B} + 2 \mathcal{A}_{,\phi}\kin + \mathcal{A}_{,\kin}\sqrt{2\kin}\mathcal{Y} ,
\eea
\ese
where $K \defn {K^{\mu}}_{\mu} = 3\hct,\mathcal{Y} \defn \dot{\kin}$ \footnote{These expressions correct two typos which are present in equations (12a) and (12b) of \cite{PearsonBattye:eos}.}.  From   (\ref{kgb_rhop}) we see that $\rho$ and $P$ have the following dependancies:
\bea
\label{eq:sec:rhop-kgb-fns}
\rho \defn \rho(\phi, \kin, K),\qquad P \defn P(\phi, \kin, \mathcal{Y}).
\eea
Since the fluid equation is $\dot{\rho} = - K(\rho+P)$, $\rho$ can only be constructed from first time derivatives of fields and so there is nothing else that $\rho$ could be a function of, while remaining at most of first order in time derivatives. The fluid equation is thus
\bea
\rho_{,\kin}\dot{\kin}   +\rho_{,\phi}\dot{\phi} + K(\rho+P +\tfrac{\dot{K}}{K}\rho_{,K} )=0.
\eea
We now want to derive the perturbed fluid variables. It is simple to use (\ref{eq:sec:rhop-kgb-fns}) to obtain $\delta\rho$ and $\delta P$ in terms of $\vphi, \delta\kin, \delta K$ and $\delta\mathcal{Y}$.
In the synchronous gauge, 
\bea
\delta K = \half \dot{h},\qquad \delta \kin = \dot{\phi}\dot{\vphi},\qquad \delta\mathcal{Y} = \ddot{\phi}\dot{\vphi} + \dot{\phi}\ddot{\vphi}.
\eea
The perturbed velocity $\theta$ and anisotropic stress $\pis$ must be computed from direct perturbation of the EMT. One finds that $\pis = 0$, and we can write the perturbed fluid variables in the form of an activation matrix,
\bea
\label{eq:sec:act-mat-kgb}
\left(\begin{array}{c} \delta - \half \frac{\rho_{,K}}{\rho}\dot{h} \\ \theta \\ \delta P \end{array}\right) = \left( \begin{array}{ccc} \frac{\rho_{,\phi}}{\rho} & \frac{\rho_{,\kin}}{\rho} & 0\\ b_1 & b_2\dot{\phi} &0 \\ P_{,\phi} &( P_{,\kin}\dot{\phi} + P_{,\mathcal{Y}}\ddot{\phi}) & P_{,\mathcal{Y}}\dot{\phi} \end{array}\right) \left(\begin{array}{c} \vphi \\ \dot{\vphi} \\ \ddot{\vphi} \end{array}\right),
\eea
where  we defined the coefficients in $\theta$'s row as
\bea
(\rho+P)b_1 \defn -  {\sqrt{2\kin}}{ } \big( \mathcal{B}_{,\kin} + 2 \mathcal{A}_{,\phi}  -   K\sqrt{2\kin} \mathcal{A}_{,\kin}\big) ,\quad (\rho+P)b_2 \defn - \mathcal{A}_{,\kin}\sqrt{2\kin}.
\eea
All components of this activation matrix are   scale independent. We have now shown that the KGB  theory has an activation matrix which is of precisely the same form as that   we derived from a model independent approach in (\ref{eq:sec:fluid-vsrs-ppf}). This means that the gauge invariant entropy perturbation is given by (\ref{eq:sec:gag-inv-entr}).

\section{Discussion}
In this paper we completed our goal of proving the claims made in our previous paper regarding the form of the equation of state for perturbations. We did this in a   model independent way, using the geometrically enlightening tensorial notation. We also showed how models with a given functional form of the Lagrangian fall into our category.

One of the clear advantages of our approach is that we are able to compute consistent cosmological perturbations in a model independent manner. Our approach provides complete   transparency as to how to relax the restrictions of reparameterization invariance or how to include more fields and/or their derivatives. However, this generality leads to a highly complicated set of equations (which we   presented in the appendices of this paper).

The result of the calculations -- equations of state for perturbations -- yields a set of modifications to the gravitational field equations which are very easy to incorporate into numerical codes, such as {\tt CAMB} \cite{Lewis:1999bs}. The modifications hold   physical significance, and, for the broad class of theories we presented in this paper, yield a small enough number of parameters that we are able to meaningfully constrain their values with current observations. This is the subject of future work.

\section*{Acknowledgements}
We have benefited from conversations with  Tessa Baker, Alex Barreira, Jolyon Bloomfield,   Pedro Ferreira, Ruth Gregory, Baojiu Li,   Adam Moss, Ian Moss, Levon Pogosian,  Ignacy Sawicki, and Costas Skordis. JAP  is supported by the STFC Consolidated Grant ST/J000426/1.
\appendix
\section{Calculation of the perturbed fluid variables}
\label{append:calc-pfvs}
Here we present   details of the calculation leading up to the activation matrix (\ref{eq:sec:fluid-vsrs-ppf}) for a general reparameterization invariant scalar-tensor theory with second order field equations. We begin by introducing some useful technology, before moving on to the explicit calculations and results.
%

\subsection{The Fourier decomposition}
There are a number of spatial derivatives within the energy-momentum tensor (EMT): we find that working in Fourier space significantly simplifies calculations, and allows tensorial notation to be maintained throughout. The advantage of this approach is that all constraints and conditions can be formulated via geometrical projections of the ``free'' tensors in the theory.

Let us begin with a space-time vector field $A_{\mu}$, whose time-like and space-like components can be explicitly isolated via $A_{\mu} = - au_{\mu} + b_{\mu}$, where $u^{\mu}b_{\mu}=0$. Then, the covariant derivative of $A_{\mu}$ is given by
\bea
\nabla_{\mu}A_{\nu} = - u_{\nu}\nabla_{\mu}a - a K_{\mu\nu} + {\gamma^{\alpha}}_{\nu}\nabla_{\mu}b_{\alpha} + b_{\alpha}{K^{\alpha}}_{\mu} u_{\nu}.
\eea
Similarly for a symmetric orthogonal space-time tensor field $B_{\mu\nu} = {\gamma^{\alpha}}_{\mu}{\gamma^{\beta}}_{\nu}B_{\alpha\beta}$,
\bea
\nabla_{\lambda}B_{\mu\nu} =  {\gamma^{\alpha}}_{\mu}{\gamma^{\beta}}_{\nu}\nabla_{\lambda}B_{\alpha\beta} + 2 {K^{\alpha}}_{\lambda}{\gamma^{\beta}}_{(\mu}u_{\nu)}B_{\alpha\beta}.
\eea
Since   this  will be useful later on, the second covariant derivative of the vector field is given by
\bea
\nabla_{\beta}\nabla_{\mu}A_{\nu} &=& - u_{\nu}\nabla_{\beta}\nabla_{\mu}a +{\gamma^{\alpha}}_{\nu} \nabla_{\beta}\nabla_{\mu}b_{\alpha} - 2K_{\nu(\mu}\nabla_{\beta)}a - a \nabla_{\beta}K_{\mu\nu} \nonumber\\
&&+ 2 {K^{(\alpha}}_{\beta}u_{\nu)} \nabla_{\mu}b_{\alpha} + u_{\nu}{K^{\alpha}}_{\mu}\nabla_{\beta}b_{\alpha} + b_{\alpha}u_{\nu}\nabla_{\beta}{K^{\alpha}}_{\mu} + {K^{\alpha}}_{\mu}K_{\nu\beta}b_{\alpha}.
\eea

We now move to Fourier space, by expanding each space-time field in Fourier modes,
\bea
B_{\mu\nu} = \int \dd^3k\, B_{(k)\mu\nu}e^{\ci kx},\qquad b_{\mu} =  \int \dd^3k\,b_{(k)\mu}e^{\ci kx},\qquad a =  \int \dd^3k\,a_{(k)}e^{\ci kx},
\eea
where $kx\defn k^{\mu}x_{\mu}$ and $k^{\mu}u_{\mu}=0$. We will always leave out the integral sign to avoid clutter. The Fourier modes are only time-dependent, and the complex exponential $e^{\ci kx}$ only has space-like derivatives,
\bea
\nabla_{\mu} e^{\ci kx} = \ci k_{\mu} e^{\ci kx}.
\eea
For example, using an obvious notation for a scalar field $F$ and its Fourier mode $F_{(k)}$, we have
\bea
\nabla_{\mu}F &=& \nabla_{\mu}(F_{(k)}e^{\ci kx})\nonumber\\
&=&\bigg[ - \dot{F}_{(k)}u_{\mu} + \ci k_{\mu}F_{(k)}\bigg] e^{\ci kx},
\eea
while for a vector field we obtain
\bea
\nabla_{\mu}A_{\nu} &=& -u_{\nu}\nabla_{\mu}(a_{(k)}e^{\ci kx}) - a_{(k)}e^{\ci kx}K_{\mu\nu} + {\gamma^{\alpha}}_{\nu}\nabla_{\mu}(b_{(k)\alpha}e^{\ci kx}) + b_{(k)\alpha}{K^{\alpha}}_{\mu}u_{\nu}e^{\ci kx}\nonumber\\
&=&\bigg[ u_{\nu}u_{\mu} \dot{a}_{(k)} - {\gamma^{\alpha}}_{\nu}u_{\mu}\dot{b}_{(k)\alpha}  - \ci k_{\mu}(u_{\nu}   a_{(k)} -{\gamma^{\alpha}}_{\nu}b_{(k)\alpha})- a_{(k)}K_{\mu\nu} + b_{(k)\alpha}{K^{\alpha}}_{\mu}u_{\nu}\bigg] e^{\ci kx},\nonumber\\
\eea
and for the orthogonal tensor field,
\bea
\nabla_{\lambda}B_{\mu\nu} &=& {\gamma^{\alpha}}_{\mu}{\gamma^{\beta}}_{\nu}\nabla_{\lambda}(B_{(k) \alpha\beta}e^{\ci kx}) + 2 {K^{\alpha}}_{\lambda}{\gamma^{\beta}}_{(\mu}u_{\nu)}B_{(k) \alpha\beta}e^{\ci kx}\nonumber\\
&=& \bigg[ -{\gamma^{\alpha}}_{\mu}{\gamma^{\beta}}_{\nu} \dot{B}_{(k)\alpha\beta}u_{\lambda} + \ci k_{\lambda}{\gamma^{\alpha}}_{\mu}{\gamma^{\beta}}_{\nu}B_{(k)\alpha\beta} + 2 {K^{\alpha}}_{\lambda}{\gamma^{\beta}}_{(\mu}u_{\nu)}B_{(k) \alpha\beta}\bigg]e^{\ci kx}.
\eea
We now proceed by evaluating some useful examples.

First, we will evaluate the Lie derivative of the metric $g_{\mu\nu}$ along the vector field $\xi_{\mu}$, given by  $\lied{\xi}g_{\mu\nu} =2\nabla_{(\mu}\xi_{\nu)}$. We parameterize the components of $\xi_{\mu}$ as $\xi_{\mu} = (-\chi u_{\mu} + {\gamma^{\nu}}_{\mu}\omega_{\nu})e^{\ci kx}$, where $\chi$ and $\omega_{\mu}$ are the Fourier modes, and we find
\bea
\lied{\xi}g_{\mu\nu}&=& 2\bigg[  \dot{\chi} u_{\mu}u_{\nu} -  ( \dot{\omega}_{\alpha}- \tfrac{1}{3}K \omega_{\alpha}  ){\gamma^{\alpha}}_{(\mu}u_{\nu)} -  \tfrac{1}{3}K \chi \gamma_{\mu\nu}-   \ci k_{(\mu}u_{\nu)}\chi +  \ci k_{(\mu} {\gamma^{\alpha}}_{\nu)}\omega_{\alpha} \bigg]e^{\ci kx}.\nonumber\\
\eea
A second useful example is evaluating the covariant derivative of the (Eulerian) perturbed metric in the synchronous gauge; the tensor field here is of the symmetric orthogonal type. Writing the Fourier mode as $\ep g_{\mu\nu} = H_{\mu\nu}e^{\ci kx}$, we find that
\bea
\label{eq:sec:nalepgdshfihsekfdsh}
\nabla_{\lambda}\ep g_{\mu\nu} =  \bigg[ -  \dot{H}_{\mu\nu}u_{\lambda} + \ci k_{\lambda} H_{\mu\nu}+\tfrac{2}{3}KH_{\lambda(\mu}u_{\nu)}\bigg]e^{\ci kx}.
\eea
Finally, the Lie derivative of the spatially isotropic energy-momentum tensor $U^{\mu\nu} = \rho u^{\mu}u^{\nu} + P\gamma^{\mu\nu}$ along $\xi^{\mu}$ is given by $\lied{\xi}U^{\mu\nu} = \xi^{\alpha}\nabla_{\alpha}U^{\mu\nu} - 2 U^{\alpha(\mu}\nabla_{\alpha}\xi^{\nu)}$, and evaluates to
\bea
\lied{\xi}U^{\mu\nu} &=& \bigg[ [2\rho\dot{\chi} - \dot{\rho}\chi]u^{\mu}u^{\nu} + 2[\ci P\chi k_{\alpha} -\rho(  \dot{\omega}_{\alpha}-\tfrac{1}{3}  K\omega_{\alpha}    )]{\gamma^{\alpha(\mu}}u^{\nu)} \nonumber\\
&&-[ \dot{P} - \tfrac{2}{3}PK]\chi\gamma^{\mu\nu} - 2 \ci Pk_{\epsilon}\omega_{\lambda}{\gamma^{\epsilon(\mu}}{\gamma^{\nu)\lambda}}\bigg]e^{\ci kx}.
\eea
For reference, the covariant derivative of $\lied{\xi}g_{\mu\nu}$ is
\bea
\nabla_{\beta}\lied{\xi} g_{\mu\nu} &=& 2 \bigg[ - u_{\beta}u_{\mu}u_{\nu}\ddot{\chi} + u_{\beta}u_{(\mu}{\gamma^{\alpha}}_{\nu)}\ddot{\omega}_{\alpha} + \bigg( K_{\mu\nu} u_{\beta}  + 2K_{\beta(\mu}u_{\nu)} + \ci k_{\beta}u_{\mu}u_{\nu} + \ci u_{\beta}k_{(\mu}u_{\nu)}\bigg)\dot{\chi}\nonumber\\
&& - \bigg( u_{\beta}{K^{\alpha}}_{(\mu}u_{\nu)} + K_{\beta(\mu}{\gamma^{\alpha}}_{\nu)} + \ci u_{\beta} k_{(\mu}{\gamma^{\alpha}}_{\nu)} + \ci k_{\beta} {\gamma^{\alpha}}_{(\mu}u_{\nu)}\bigg)\dot{\omega}_{\alpha}\nonumber\\
&& - \bigg( \nabla_{\beta}K_{\mu\nu} - k_{\beta} k_{(\mu}u_{\nu)}  + \ci k_{\beta}K_{\mu\nu} + \ci K_{\beta(\mu}k_{\nu)}\bigg)\chi\nonumber\\
&& + \bigg( {K^{\alpha}}_{(\mu}K_{\nu)\beta} + \ci k_{\beta}{K^{\alpha}}_{(\mu}u_{\nu)} - k_{\beta}{\gamma^{\alpha}}_{(\mu}k_{\nu)}  + \tfrac{1}{2}u_{\mu}\nabla_{\beta}{K^{\alpha}}_{\nu} + \tfrac{1}{2}u_{\nu}\nabla_{\beta}{K^{\alpha}}_{\mu}\bigg)\omega_{\alpha}\bigg] e^{\ci kx}.\nonumber\\
\eea
\subsection{Perturbed EMT}
Here we lay out the Fourier decomposition of the perturbed EMT. This is performed by writing
\bea
\ep {U^{\mu}}_{\nu} = \delta\rho u^{\mu}u_{\nu} + (\rho+P)v^{(\mu}u_{\nu)} + \delta P {\gamma^{\mu}}_{\nu} + P{\Pi^{\mu}}_{\nu}.
\eea
Note that the mixed EMT is obtained from the contravariant EMT via
\bea
\label{eq:sec;mixed-from-contra}
\ep {U^{\mu}}_{\nu} = g_{\alpha\nu}\ep U^{\mu\alpha} + U^{\alpha\mu}\ep g_{\nu\alpha}.
\eea
The perturbed fluid variables can be obtained from $\ep U^{\mu\nu} $ by application of various ``projectors'',
\bse
\label{eq:sec:projectors}
\bea
\delta \rho &=& u_{\mu}u_{\nu}\ep U^{\mu\nu} ,\\
v^{\alpha} &=& - \tfrac{1}{\rho+P} u_{\mu}{\gamma_{\nu}}^{\alpha}\ep U^{\mu\nu} ,\\
\delta P &=& \tfrac{1}{3}\gamma_{\mu\nu}\ep U^{\mu\nu} ,\\
P\Pi^{\alpha\beta} &=&{\perp^{ \alpha\beta}}_{ \mu\nu} \ep U^{\mu\nu} .
\eea
\ese
We made use of the transverse-traceless orthogonal projection operator, ${\perp^{ \alpha\beta}}_{ \mu\nu}$, defined in (\ref{eq:sec:defn-perp}).

We note that the Lie derivatives are given by
\bea
\lied{\xi}\phi = \xi^{\alpha}\nabla_{\alpha}\phi,\quad \lied{\xi}g_{\mu\nu} = 2\nabla_{(\mu}\xi_{\nu)},\quad \lied{\xi}U^{\mu\nu} = \xi^{\alpha}\nabla_{\alpha}U^{\mu\nu} - 2 U^{\alpha(\mu}\nabla_{\alpha}\xi^{\nu)}.
\eea

The Fourier decompositions of the scalar field $\ep\phi$, gauge field $\xi^{\mu}$ and metric perturbation  are given by
\bse
\label{eq:sec:four-decomp-sdefinitiions}
\bea
&\ep \phi = \vphi  e^{\ci kx},\quad \xi_{\mu} = \zeta_{\mu}e^{\ci kx},\quad \ep g_{\mu\nu} = H_{\mu\nu}e^{\ci kx},&
\\
&v_{\mu} = V_{\mu}e^{\ci kx},\quad \Pi_{\mu\nu} = \pi_{\mu\nu}e^{\ci kx}.&
\eea
\ese
where it is to be understood that $\{\vphi,\zeta_{\mu}, H_{\mu\nu}, V_{\mu}, \pi_{\mu\nu}\}$ are the Fourier modes (and as such, only have time-like derivatives). In the synchronous gauge, the metric perturbation is only space-like, and so satisifies
\bea
H_{\mu\nu} = {\gamma^{\alpha}}_{\mu}{\gamma^{\beta}}_{\nu}H_{\alpha\beta}.
\eea
We   isolate the time-like and space-like parts of the Fourier mode $\zeta_{\mu}$ of the gauge field $\xi_{\mu}$ by writing
\bea
\zeta_{\mu} = - u_{\mu}\chi +\omega_{\mu},
\eea
where
\bea
\omega_{\mu} =  {\gamma^{\alpha}}_{\mu}\omega_{\alpha}.
\eea
We will decompose the space-like vector $k_{\mu}$ into a scalar $k$ multiplying a unit space-like vector $\hat{k}_{\mu}$ via
\bea
\label{scalk}
k_{\mu} = \ci k \hat{k}_{\mu},\qquad \hat{k}_{\mu} = {\gamma^{\nu}}_{\mu}\hat{k}_{\nu},\qquad \hat{k}^{\mu}\hat{k}_{\mu}=1.
\eea
When needed, we will decompose the Fourier modes $H_{\mu\nu}, \omega_{\alpha}, V_{\mu}$ and $\pi_{\mu\nu}$ into scalars via
\bse
\bea
\label{eq:sec:h-defn-hl-ht}
H_{ \alpha\beta} &=& \tfrac{1}{3} \gamma_{\alpha\beta}\qsubrm{H}{L} + \hat{k}_{\rho}\hat{k}_{\sigma}{\perp^{\rho\sigma}}_{\alpha\beta}\qsubrm{H}{T},\\
\omega_{\mu} &=& \qsuprm{\omega}{s}k_{\mu},\\
V_{\mu} &=&-k \tis \hat{k}_{\mu},\\
\pi_{\alpha\beta} &=& \hat{k}_{\rho}\hat{k}_{\sigma}{\perp^{\rho\sigma}}_{\alpha\beta} \pis
\eea
\ese
We reiterate that we are interested in scalar perturbations in this paper.

We will relate the longitudinal, $\qsubrm{H}{L}$, and transverse, $\qsubrm{H}{T}$, modes of the metric perturbation to the synchronous gauge variables $h$ and $\eta$. Note that
\bse
\bea
&k(\rho+P)\tis =\hat{k}_{\mu}u_{\nu}\ep U^{\mu\nu},&\\
& \gamma^{\alpha\beta}H_{\alpha\beta} =   \qsubrm{H}{L},\qquad {\perp^{\mu\nu}}_{\alpha\beta}H_{\mu\nu} = \hat{k}_{\rho}\hat{k}_{\sigma}{\perp^{\rho\sigma}}_{\alpha\beta}\qsubrm{H}{T}.&
\eea
\ese
By using (\ref{eq:sec;mixed-from-contra}), in the synchronous gauge the mixed EMT is deduced via
\bea
\label{eq:sec:mixed-emt-withP}
\ep {U^{\mu}}_{\nu}&=& U^{\mu\alpha}\ep g_{\nu\alpha} + g_{\nu\alpha}\ep U^{\mu\alpha}\nonumber\\
&=& \tfrac{1}{3}P{\gamma^{\mu}}_{\nu} \qsubrm{H}{L}+ P\hat{k}_{\rho}\hat{k}_{\sigma}  {\perp^{\rho\sigma\mu}}_{\nu }\qsubrm{H}{T} + g_{\nu\alpha}\ep U^{\mu\alpha}.
\eea

\subsection{Evaluating the perturbed EMT}

We now use this technology to derive the activation matrix for the scalar field theory described in section \ref{sec:schem_summ_perfvars}. Using geometric projectors we will be able to isolate   the tensors and   their scale dependence which multiply the field variables that are used to construct the perturbed fluid variables; we will also be able to impose reparameterization invariance at the tensorial level.  

From the Lagrangian for perturbations, the Eulerian perturbed EMT is computed via
\bea
\label{eq:sec:ep-emt-fdljfhdkshfdkh}
\ep U^{\mu\nu} = \hat{\mathbb{Y}}^{\mu\nu}\ep\phi + \hat{\mathbb{W}}^{\mu\nu\alpha\beta} \ep g_{\alpha\beta} + \hat{\mathbb{Y}}^{\mu\nu}\lied{\xi}\phi + \hat{\mathbb{W}}^{\mu\nu\alpha\beta} \lied{\xi}g_{\alpha\beta} - \lied{\xi}U^{\mu\nu}.
\eea
Note that we have included the gauge field $\xi_{\mu}$ explicitly: when the parameters in the theory are arranged to make it deouple, the theory is reparameterization invariant. 

For the Eulerian perturbed EMT   (\ref{eq:sec:ep-emt-fdljfhdkshfdkh}), we take the derivative operators to be
\bse
\label{eq:sec;oprators-yw-exp-defn}
\bea
\hat{\mathbb{Y}}^{\mu\nu} &=& \mathbb{A}^{\mu\nu} + \mathbb{B}^{\alpha\mu\nu}\nabla_{\alpha} + \mathbb{C}^{\alpha\beta\mu\nu}\nabla_{\alpha}\nabla_{\beta} + \mathbb{D}^{\rho\alpha\beta\mu\nu} \nabla_{\rho}\nabla_{\alpha}\nabla_{\beta},\\
\hat{\mathbb{W}}^{\mu\nu\alpha\beta} &=& \mathbb{E}^{\mu\nu\alpha\beta} + \mathbb{F}^{\rho\mu\nu\alpha\beta}\nabla_{\rho};
\eea
\ese
that is, we have set $\mathbb{G}=0$.
We will write the Eulerian perturbed scalar field and the Eulerian perturbed metric as in (\ref{eq:sec:four-decomp-sdefinitiions}).
After using the Fourier decompositions, the EMT is given by
\bea
\label{eq:sec:emt-letter-ex}
e^{-\ci kx} \ep U^{\mu\nu} &=& \mathsf{Y}^{(0)\mu\nu}\vphi+  \mathsf{Y}^{(1)\mu\nu}\dot{ \vphi}+  \mathsf{Y}^{(2)\mu\nu}\ddot{ \vphi}+\mathsf{Y}^{(3)\mu\nu}\dddot{ \vphi}\nonumber\\
&&+ \mathsf{W}^{(0)\mu\nu\alpha\beta}H_{ \alpha\beta}+  \mathsf{W}^{(1)\mu\nu\alpha\beta}\dot{H}_{ \alpha\beta} + e^{-\ci kx}\Delta_{\{\xi\}} \ep U^{\mu\nu},
\eea
where $ \Delta_{\{\xi\}} \ep U^{\mu\nu}$ is the contribution to the EMT from the gauge field,  given by
\bea
\label{eq:sec:gag-fld-contrib}
\Delta_{\{\xi\}} \ep U^{\mu\nu}  &=& \hat{\mathbb{W}}^{ \rho\sigma\mu\nu}\lied{\xi}g_{\mu\nu} +\hat{\mathbb{Y}}^{ \rho\sigma  }\lied{\xi}\phi - \lied{\xi}U^{\mu\nu} .
\eea

The coefficients of each time derivative of $\vphi$ and $H_{\mu\nu}$ in (\ref{eq:sec:emt-letter-ex}) are given by
\bse
\bea
 \mathsf{Y}^{(0)\mu\nu} &=&  \mathbb{A}^{\mu\nu} +\ci k_{\alpha} \mathbb{B}^{\alpha\mu\nu}- k_{\alpha}k_{\beta} \mathbb{C}^{\alpha\beta\mu\nu}  - \ci k_{\rho}k_{\alpha}k_{\beta}\mathbb{D}^{\rho\alpha\beta\mu\nu}  ,\\
 \mathsf{Y}^{(1)\mu\nu} &=& - u_{\alpha}\mathbb{B}^{\alpha\mu\nu} - K_{\alpha\beta}\mathbb{C}^{\alpha\beta\mu\nu}  \nonumber\\
 && -\ci k_{\epsilon}[  2  {\gamma^{\epsilon}}_{(\alpha} u_{\beta)}\mathbb{C}^{\alpha\beta\mu\nu}+ K_{\rho\alpha}\mathbb{D}^{\rho\alpha\epsilon\mu\nu}+   K_{\alpha\beta}(\mathbb{D}^{\alpha\epsilon\beta\mu\nu}+  \mathbb{D}^{\epsilon\alpha\beta\mu\nu})]\nonumber\\
 &&  + u_{\lambda}k_{\rho}k_{\epsilon} \mathbb{D}^{\lambda\epsilon\rho\mu\nu},\\
\mathsf{Y}^{(2)\mu\nu} &=& u_{\alpha}u_{\beta} \mathbb{C}^{\alpha\beta\mu\nu}  + \tfrac{1}{3}K(u_{\rho}\gamma_{\alpha\beta} +2\gamma_{\rho(\alpha}u_{\beta)} )\mathbb{D}^{\rho\alpha\beta\mu\nu}\nonumber\\
&&    +  \ci k_{\epsilon}(  u_{\alpha}u_{\beta} \mathbb{D}^{\epsilon\alpha\beta\mu\nu}+ 2   u_{\rho}{\gamma^{\epsilon}}_{(\alpha}u_{\beta)}\mathbb{D}^{\rho\alpha\beta\mu\nu}),\\
\label{eq:y3-threetimederivslpphi}
\mathsf{Y}^{(3)\mu\nu} &=& - u_{\rho}u_{\alpha}u_{\beta}\mathbb{D}^{\rho\alpha\beta\mu\nu},\\
\mathsf{W}^{(0)\mu\nu\alpha\beta} &=&  \mathbb{E}^{\mu\nu\alpha\beta}  +\tfrac{2}{3}K{\gamma^{\alpha}}_{\rho}{\gamma^{\beta}}_{(\pi}u_{\epsilon)}\mathbb{F}^{\rho\mu\nu\pi\epsilon}+ \ci k_{\rho}\mathbb{F}^{\rho\mu\nu\alpha\beta} ,\\
\mathsf{W}^{(1)\mu\nu\alpha\beta} &=& - u_{\rho}\mathbb{F}^{\rho\mu\nu\alpha\beta} .
\eea
\ese
Before we proceed any further, notice that (\ref{eq:y3-threetimederivslpphi}) represents the contribution of third time derivatives of $\vphi$ to the perturbed energy-momentum tensor (\ref{eq:sec:emt-letter-ex}). These terms are clearly problematic, but can be remedied by setting $u_{\rho}u_{\alpha}u_{\beta}\mathbb{D}^{\rho\alpha\beta\mu\nu}=0$. Thus, we take $\mathsf{Y}^{(3)\mu\nu}=0$ in everything that follows.
We have written these in such a way that the individual terms are grouped in order of scale, $k_{\mu}$.  We now use (\ref{scalk}) to explicitly isolate the scale dependence, yielding
\bse
\label{eq:sectimdcoeffs-wy}
\bea
\mathsf{Y}^{(0)\mu\nu}  &=& \mathsf{Y}^{(0,0)\mu\nu} +\mathsf{Y}^{(0,1)\mu\nu}k +\mathsf{Y}^{(0,2)\mu\nu}k^2 +\mathsf{Y}^{(0,3)\mu\nu} k^3,\\
\mathsf{Y}^{(1)\mu\nu}  &=& \mathsf{Y}^{(1,0)\mu\nu} +\mathsf{Y}^{(1,1)\mu\nu}k +\mathsf{Y}^{(1,2)\mu\nu}k^2  ,\\
\mathsf{Y}^{(2)\mu\nu}  &=& \mathsf{Y}^{(2,0)\mu\nu} +\mathsf{Y}^{(2,1)\mu\nu}k ,\\
\mathsf{W}^{(0)\mu\nu\alpha\beta} &=& \mathsf{W}^{(0,0)\mu\nu\alpha\beta} +\mathsf{W}^{(0,1)\mu\nu\alpha\beta} k,\\
\label{eq:sec:w1-w10}
\mathsf{W}^{(1)\mu\nu\alpha\beta} &=&\mathsf{W}^{(1,0)\mu\nu\alpha\beta} .
\eea
\ese
A glance at (\ref{eq:sec:w1-w10})  shows that     any coefficient of $\dot{h}$ or $\dot{\eta}$ which may be present will always be scale independent. Also, there is no $k^2$ dependence of any coefficients of $\ddot{\vphi}$ (this is evident from the lack of a $ \mathsf{Y}^{(2,2)\mu\nu}$-term). 
The time-dependent coefficients of each term in (\ref{eq:sectimdcoeffs-wy}) are given by
\bse
\bea
\label{eq:sec:Y00-A}
\mathsf{Y}^{(0,0)\mu\nu}  &\defn& \mathbb{A}^{\mu\nu} ,\\
\mathsf{Y}^{(0,1)\mu\nu}  &\defn& -\hat{k}_{\alpha} \mathbb{B}^{\alpha\mu\nu},\\
\mathsf{Y}^{(0,2)\mu\nu}  &\defn& \hat{k}_{\alpha}\hat{k}_{\beta} \mathbb{C}^{\alpha\beta\mu\nu},\\
\mathsf{Y}^{(0,3)\mu\nu}  &\defn& \hat{k}_{\rho}\hat{k}_{\alpha}\hat{k}_{\beta}\mathbb{D}^{\rho\alpha\beta\mu\nu},
\eea
\bea
\label{eq:sec:Y10-fdjhfdj}
 \mathsf{Y}^{(1,0)\mu\nu} &\defn& - u_{\alpha}\mathbb{B}^{\alpha\mu\nu} - K_{\alpha\beta}\mathbb{C}^{\alpha\beta\mu\nu}-( \nabla_{\rho}K_{\alpha\beta})\mathbb{D}^{\rho\alpha\beta\mu\nu},\\
 \mathsf{Y}^{(1,1)\mu\nu} &\defn& \hat{k}_{\epsilon}[ 2  {\gamma^{\epsilon}}_{(\alpha} u_{\beta)}\mathbb{C}^{\alpha\beta\mu\nu}+ K_{\rho\alpha}\mathbb{D}^{\rho\alpha\epsilon\mu\nu}+   K_{\rho\beta}\mathbb{D}^{\rho\epsilon\beta\mu\nu}+  K_{\alpha\beta}\mathbb{D}^{\epsilon\alpha\beta\mu\nu}],\\
 \mathsf{Y}^{(1,2)\mu\nu} &\defn&- u_{\lambda} \hat{k}_{\epsilon}\hat{k}_{\rho} \mathbb{D}^{\lambda\epsilon\rho\mu\nu},
 \eea
 \bea
 \label{eq:sec:Y20}
  \mathsf{Y}^{(2,0)\mu\nu} &\defn&  u_{\alpha}u_{\beta} \mathbb{C}^{\alpha\beta\mu\nu}  + \tfrac{1}{3}K(u_{\rho}\gamma_{\alpha\beta} +2\gamma_{\rho(\alpha}u_{\beta)} )\mathbb{D}^{\rho\alpha\beta\mu\nu},\\
  \mathsf{Y}^{(2,1)\mu\nu} &\defn& -  \hat{k}_{\epsilon}(  u_{\alpha}u_{\beta} \mathbb{D}^{\epsilon\alpha\beta\mu\nu}+ 2   u_{\rho}{\gamma^{\epsilon}}_{(\alpha}u_{\beta)}\mathbb{D}^{\rho\alpha\beta\mu\nu}),
  \eea
  \bea
\label{eq:sec:W00}
\mathsf{W}^{(0,0)\mu\nu\alpha\beta} &\defn&\mathbb{E}^{\mu\nu\alpha\beta} +\tfrac{2}{3}K{\gamma^{\alpha}}_{\rho}{\gamma^{\beta}}_{(\pi}u_{\epsilon)}\mathbb{F}^{\rho\mu\nu\pi\epsilon},\\
\label{w01-uF}
\mathsf{W}^{(0,1)\mu\nu\alpha\beta} &\defn& -\hat{k}_{\rho}\mathbb{F}^{\rho\mu\nu\alpha\beta},\\
\label{w10-uF}
\mathsf{W}^{(1,0)\mu\nu\alpha\beta} &\defn& u_{\rho}\mathbb{F}^{\rho\mu\nu\alpha\beta} .
\eea
\ese
Just to explain the labels on these objects (we will be introducing another set later on when we look at the gauge fields influence on the system): $\mathsf{Y}^{(\rm X,Y)\mu\nu}$ is the coefficient of the $\qsuprm{\rm X}{th}$-time derivative and $\qsuprm{\rm Y}{th}$ multiple of $k$ infront of $\vphi$; these coefficients   explicitly only have time dependence. An obvious extension to the $\mathsf{W}^{(\rm X,Y)\mu\nu\alpha\beta}$ as time dependent coefficients  of $H_{\alpha\beta}$.   There is also a nice structure which emerges:
\bse
\bea
\mathsf{Y}^{(\rm N)\mu\nu} &=& \sum_{\rm n=0}^{3-\rm N}\mathsf{Y}^{(\rm N,n)\mu\nu}k^{\rm n},\qquad \rm N = 0,1,2,\\
\mathsf{W}^{(\rm N)\mu\nu\alpha\beta} &=& \sum_{\rm n=0}^{1-\rm N}\mathsf{W}^{(\rm N,n)\mu\nu\alpha\beta}k^{\rm n},\qquad \rm N = 0,1.
\eea
\ese
The upper limits in these sums are set by the number of derivatives we used in the operator expansions of $\hat{\mathbb{Y}}^{\mu\nu}, \hat{\mathbb{W}}^{\mu\nu\alpha\beta}$. 

 In a similar fashion, the gauge field contribution (\ref{eq:sec:gag-fld-contrib}) can be written as
\bea
 e^{-\ci kx}\Delta_{\{\xi\}} \ep U^{ \rho\sigma} &=&\Theta^{(0)\rho\sigma}\chi + \Theta^{(1)\rho\sigma}\dot{\chi} + \Theta^{(2)\rho\sigma}\ddot{\chi} \nonumber\\
&& + \Xi^{(0)\rho\sigma\alpha}\omega_{\alpha}+ \Xi^{(1)\rho\sigma\alpha}\dot{\omega}_{\alpha}+ \Xi^{(2)\rho\sigma\alpha}\ddot{\omega}_{\alpha},
\eea
where the coefficient of each time derivative is
\bse
\bea
\Theta^{(0)\rho\sigma}&\defn& -  \mathsf{Y}^{(0) \rho\sigma}   \dot{\phi}  -\mathsf{Y}^{(1) \rho\sigma}   \ddot{\phi} -  \mathsf{Y}^{(2) \rho\sigma} \dddot{\phi} +\dot{\rho}u^{\rho}u^{\sigma}+[ \dot{P} - \tfrac{2}{3}PK] \gamma^{\rho\sigma} \nonumber\\
&&-2 \big[ \nabla_{\beta}K_{\mu\nu}   \mathbb{F}^{\beta\rho\sigma\mu\nu}+ \tfrac{1}{3}K   \gamma_{\mu\nu}\mathbb{E}^{\rho\sigma\mu\nu}- k_{\beta} k_{(\mu}u_{\nu)}   \mathbb{F}^{\beta\rho\sigma\mu\nu} \nonumber\\
&&+ \ci k_{\epsilon}( P{\gamma^{\epsilon(\rho}}u^{\sigma)}   + K_{\mu\nu}  \mathbb{F}^{\epsilon\rho\sigma\mu\nu}+  K_{\alpha(\mu} {\gamma^{\epsilon}}_{\beta)} \mathbb{F}^{\alpha\rho\sigma\mu\beta}+ {\gamma^{\epsilon}}_{(\beta}u_{\mu)}\mathbb{E}^{\rho\sigma\mu\beta})\big],\\
\Theta^{(1)\rho\sigma} &\defn & -\mathsf{Y}^{(1) \rho\sigma} \dot{\phi} - 2 \mathsf{Y}^{(2) \rho\sigma}\ddot{\phi} +2\big[ -K_{\mu\nu}    \mathsf{W}^{(1)\rho\sigma\mu\nu} + 2K_{\beta(\mu}u_{\nu)}   \mathbb{F}^{\beta\rho\sigma\mu\nu}\nonumber\\
&&+  u_{\mu}u_{\nu}\mathbb{E}^{\rho\sigma\mu\nu}+ \ci k_{\beta}(u_{\mu}u_{\nu}  \mathbb{F}^{\beta\rho\sigma\mu\nu}-    {\gamma^{\beta}}_{(\mu}u_{\nu)}  \mathsf{W}^{(1)\rho\sigma\mu\nu})\big]-2\rho u^{\rho}u^{\sigma} ,\\
\Theta^{(2)\rho\sigma} &\defn &  2 u_{\mu}u_{\nu}  \mathsf{W}^{(1)\rho\sigma\mu\nu}
  - \mathsf{Y}^{(2) \rho\sigma} \dot{\phi} ,
\eea
\bea
\Xi^{(0)\rho\sigma\alpha}&\defn& 2 \big[ \tfrac{1}{3}  K( -\rho{\gamma^{\alpha(\rho}}u^{\sigma)}+ {\gamma^{\alpha}}_{(\mu}u_{\nu)} \mathbb{E}^{\rho\sigma\mu\nu} )+({K^{\alpha}}_{(\mu}K_{\nu)\beta}    + \tfrac{1}{2}u_{\mu}\nabla_{\beta}{K^{\alpha}}_{\nu} ) \mathbb{F}^{\beta\rho\sigma\mu\nu}\nonumber\\
&&+ \ci k_{\epsilon}({K^{\alpha}}_{(\mu}u_{\nu)}   \mathbb{F}^{\epsilon\rho\sigma\mu\nu} +     {\gamma^{\epsilon}}_{(\mu}{\gamma^{\alpha}}_{\nu)}\mathbb{E}^{\rho\sigma\mu\nu}  +    P {\gamma^{\epsilon(\rho}}{\gamma^{\sigma)\alpha}})- k_{\beta} k_{(\mu} {\gamma^{\alpha}}_{\nu)} \mathbb{F}^{\beta\rho\sigma\mu\nu}\big]  ,\nonumber\\ \\
\Xi^{(1)\rho\sigma\alpha}&\defn&  -2 \big[ -\rho  {\gamma^{\alpha(\rho}}u^{\sigma)}+    {\gamma^{\alpha}}_{(\mu}u_{\nu)} \mathbb{E}^{\rho\sigma\mu\nu} + K_{\beta(\mu}{\gamma^{\alpha}}_{\nu)}   \mathbb{F}^{\beta\rho\sigma\mu\nu} -{K^{\alpha}}_{(\mu}u_{\nu)}  \mathsf{W}^{(1)\rho\sigma\mu\nu}\nonumber\\
&& +  \ci k_{\beta}( {\gamma^{\alpha}}_{(\mu}u_{\nu)}  \mathbb{F}^{\beta\rho\sigma\mu\nu} -     {\gamma^{\beta}}_{(\mu}{\gamma^{\alpha}}_{\nu)}  \mathsf{W}^{(1)\rho\sigma\mu\nu}) \big] ,\\
\Xi^{(2)\rho\sigma\alpha}&\defn&-2 u_{(\mu}{\gamma^{\alpha}}_{\nu)}  \mathsf{W}^{(1)\rho\sigma\mu\nu}.
\eea
\ese
Explicitly isolating the scale dependence of these expressions yields
\bse
\label{eq:sec;tdcs-theta-xi}
\bea
\Theta^{(0)\rho\sigma} &=&\Theta^{(0,0)\rho\sigma} +\Theta^{(0,1)\rho\sigma}k + \Theta^{(0,2)\rho\sigma}k^2+ \Theta^{(0,3)\rho\sigma}k^3,\\
\Theta^{(1)\rho\sigma} &=&\Theta^{(1,0)\rho\sigma} +\Theta^{(1,1)\rho\sigma}k + \Theta^{(1,2)\rho\sigma}k^2,\\
\Theta^{(2)\rho\sigma} &=&\Theta^{(2,0)\rho\sigma} +\Theta^{(2,1)\rho\sigma}k  ,\\
\Xi^{(0)\rho\sigma\alpha} &=& \Xi^{(0,0)\rho\sigma\alpha}+\Xi^{(0,1)\rho\sigma\alpha}k + \Xi^{(0,2)\rho\sigma\alpha}k^2,\\
\Xi^{(1)\rho\sigma\alpha}&=&\Xi^{(1,0)\rho\sigma\alpha}+\Xi^{(1,1)\rho\sigma\alpha}k,\\
\Xi^{(2)\rho\sigma\alpha} &=& \Xi^{(2,0)\rho\sigma\alpha}.
\eea
\ese
The time-dependent coefficients in (\ref{eq:sec;tdcs-theta-xi}) are
\bse
\label{eq:Sectcds--b43}
\bea
\label{eq:sec:theta00-sosftq}
\Theta^{(0,0)\rho\sigma} & \defn & - \mathsf{Y}^{(0,0)\rho\sigma}\dot{\phi}  - \mathsf{Y}^{(1,0)\rho\sigma}\ddot{\phi} - \mathsf{Y}^{(2,0)\rho\sigma}\dddot{\phi}+\dot{\rho}u^{\rho}u^{\sigma}+[ \dot{P} - \tfrac{2}{3}PK] \gamma^{\rho\sigma} \nonumber\\
&&+\tfrac{2}{3} \big[ \dot{K} \gamma_{\mu\nu}  \mathsf{W}^{(1,0)\rho\sigma\mu\nu}- K( \gamma_{\mu\nu}\mathbb{E}^{\rho\sigma\mu\nu}+\tfrac{2}{3}K\gamma_{\beta(\mu}u_{\nu)} \mathbb{F}^{\beta\rho\sigma\mu\nu}       ) \big],\\
\Theta^{(0,1)\rho\sigma} & \defn & - \mathsf{Y}^{(0,1)\rho\sigma}\dot{\phi}  - \mathsf{Y}^{(1,1)\rho\sigma}\ddot{\phi} +\Theta^{(2,1)\rho\sigma}\tfrac{\dddot{\phi}}{\dot{\phi}}-2( (\rho+P)\hat{k}_{\epsilon}{\gamma^{\epsilon(\rho}}u^{\sigma)} \nonumber\\
&&  -\tfrac{1}{3} K\gamma_{\mu\nu}   \mathsf{W}^{(0,1)\rho\sigma\mu\nu} - \tfrac{1}{2}\hat{k}_{\alpha}\Xi^{(1,0)\rho\sigma\alpha} - \tfrac{1}{6}K\hat{k}_{\alpha}\Xi^{(2,0)\rho\sigma\alpha}  )  ,\\
\Theta^{(0,2)\rho\sigma} & \defn & - \mathsf{Y}^{(0,2)\rho\sigma}\dot{\phi} +\Theta^{(1,2)\rho\sigma}\tfrac{\ddot{\phi}}{\dot{\phi}} -  \hat{k}_{\alpha}\Xi^{(2,0)\rho\sigma\alpha},\\
\Theta^{(0,3)\rho\sigma} & \defn & - \mathsf{Y}^{(0,3)\rho\sigma}\dddot{\phi}  ,
\eea
\bea
\Theta^{(1,0)\rho\sigma} &\defn& -\mathsf{Y}^{(1,0) \rho\sigma} \dot{\phi} - 2 \mathsf{Y}^{(2,0) \rho\sigma}\ddot{\phi} -2\rho u^{\rho}u^{\sigma} \nonumber\\
&&+2\big[ -\tfrac{1}{3}K\gamma_{\mu\nu}    \mathsf{W}^{(1,0)\rho\sigma\mu\nu} + \tfrac{2}{3}K\gamma_{\beta(\mu}u_{\nu)}   \mathbb{F}^{\beta\rho\sigma\mu\nu} +  u_{\mu}u_{\nu}\mathbb{E}^{\rho\sigma\mu\nu} \big],\\
\Theta^{(1,1)\rho\sigma} &\defn& -\mathsf{Y}^{(1,1) \rho\sigma} \dot{\phi}+2u_{\mu}u_{\nu}  \mathsf{W}^{(0,1)\rho\sigma\mu\nu}-   \hat{k}_{\alpha}\Xi^{(2,0)\rho\sigma\alpha} + 2\Theta^{(2,1)\rho\sigma}\tfrac{\ddot{\phi}}{\dot{\phi}}  ,
\eea
\bea
\Theta^{(1,2)\rho\sigma} &\defn& -\mathsf{Y}^{(1,2) \rho\sigma} \dot{\phi}     ,\\
\Theta^{(2,0)\rho\sigma} &\defn&2 u_{\mu}u_{\nu}  \mathsf{W}^{(1,0)\rho\sigma\mu\nu}
  - \mathsf{Y}^{(2,0) \rho\sigma} \dot{\phi} ,\\
\Theta^{(2,1)\rho\sigma} &\defn&   - \mathsf{Y}^{(2,1) \rho\sigma} \dot{\phi} ,
\eea
\bea
\Xi^{(0,0)\rho\sigma\alpha}&\defn &    \tfrac{2}{3}  K( -\rho{\gamma^{\alpha(\rho}}u^{\sigma)}+ {\gamma^{\alpha}}_{(\mu}u_{\nu)} \mathbb{E}^{\rho\sigma\mu\nu}+\tfrac{1}{3}K{\gamma^{\alpha}}_{(\mu}\gamma_{\nu)\beta}\mathbb{F}^{\beta\rho\sigma\mu\nu})\nonumber\\
&&+ \tfrac{1}{6}\Xi^{(2,0)\rho\sigma\alpha}\dot{K}  + \tfrac{1}{3}K \tfrac{1}{3}K(u_{\mu}{\gamma^{\alpha}}_{\beta}u_{\nu}\mathbb{F}^{\beta\rho\sigma\mu\nu} +  u_{\mu}{\gamma}_{\nu\beta}u^{\alpha}\mathbb{F}^{\beta\rho\sigma\mu\nu} ),\\
\label{eq:sec:xi01}
\Xi^{(0,1)\rho\sigma\alpha}&\defn & 2\hat{k}_{\epsilon}(\tfrac{1}{3}K{\gamma^{\alpha}}_{(\mu}u_{\nu)}    \mathbb{F}^{\epsilon\rho\sigma\mu\nu} +   {\gamma^{\epsilon}}_{(\mu}{\gamma^{\alpha}}_{\nu)}\mathbb{E}^{\rho\sigma\mu\nu}  +    P {\gamma^{\epsilon(\rho}}{\gamma^{\sigma)\alpha}}),\\
\label{eq:sec:xi02}
\Xi^{(0,2)\rho\sigma\alpha}&\defn& -2  \hat{k}_{\epsilon}{\gamma^{\epsilon}}_{(\mu} {\gamma^{\alpha}}_{\nu)} \mathsf{W}^{(0,1)\rho\sigma\mu\nu},
\eea
\bea
\Xi^{(1,0)\rho\sigma\alpha} &\defn& -2 \big[ -\rho  {\gamma^{\alpha(\rho}}u^{\sigma)}+    {\gamma^{\alpha}}_{(\mu}u_{\nu)} \mathbb{E}^{\rho\sigma\mu\nu} \nonumber\\
&&+ \tfrac{1}{3}K\gamma_{\beta(\mu}{\gamma^{\alpha}}_{\nu)}   \mathbb{F}^{\beta\rho\sigma\mu\nu} +\tfrac{1}{6}K\Xi^{(2,0)\rho\sigma\alpha}\big] ,\\
\label{eq:ec;xi11}
\Xi^{(1,1)\rho\sigma\alpha} &\defn&  2 ( {\gamma^{\alpha}}_{(\mu}u_{\nu)}   \mathsf{W}^{(0,1)\rho\sigma\mu\nu}+     \hat{k}_{\epsilon}{\gamma^{\epsilon}}_{(\mu}{\gamma^{\alpha}}_{\nu)}  \mathsf{W}^{(1,0)\rho\sigma\mu\nu}) ,
\eea
\bea
\label{eq:sec:xi20}
\Xi^{(2,0)\rho\sigma\alpha}&\defn&-2 u_{(\mu}{\gamma^{\alpha}}_{\nu)}  \mathsf{W}^{(1,0)\rho\sigma\mu\nu}.
\eea
\ese
These satisfy the structure
\bse
\bea
\Theta^{(\rm N)\rho\sigma}  &=& \sum_{\rm N=0}^{3-\rm N}\Theta^{(\rm N, n)\rho\sigma} k^{\rm n},\qquad {\rm N} = 0,1,2,\\
\Xi^{(\rm N)\rho\sigma\alpha}  &=& \sum_{\rm N=0}^{2-\rm N}\Xi^{(\rm N, n)\rho\sigma\alpha} k^{\rm n},\qquad {\rm N} = 0,1,2.
\eea
\ese

We have now obtained all the ``basic'' equations required. To recap what we have done: we have used purely geometrical projectors to isolate the scale and time dependence of the extra  fields $\vphi, H_{\mu\nu}$ (these are the Fourier modes of the scalar field perturbation $\ep\phi$ and metric perturbation $\ep g_{\mu\nu}$) which appear in a dark sector theory. We have also isolated how the gauge field components $\chi \sim \xi^0$ and $\omega^i \sim \xi^i$ (where ``$\sim$'' denotes that the field is the Fourier mode) enter into the perturbed energy-momentum tensor, again, isolating how each time and space derivative enters.

It is worth pointing out again that all tensors in the EMT   can be traced back to an effective Lagrangian for perturbation. 

We   now show how to impose two   important theoretical priors upon the theory: (i) second order field equations and (ii) reparameterization invariance. Between (i) and (ii) we will show how the ``naive'' activation matrix can be obtained -- it is naive in the sense that reparamterization invariance has not yet been imposed on the components of the matrix.

\subsection{Second order field equation constraints}
The conditions for second order field equations are obtained by removing time derivatives of all fields of order 2 and above from expressions for $\delta\rho$ and $\tis$. This amounts to requiring
\bea
\label{eq:sec:conds-sofeqs}
u_{\mu}u_{\nu}\mathsf{Y}^{(2)\mu\nu}=0,\qquad u_{\mu}{\gamma^{\alpha}}_{\nu}\mathsf{Y}^{(2)\mu\nu}=0.
\eea
\subsection{Naive fluid variables}
Here we show which projections of the tensors (\ref{eq:sec:emt-letter-ex}) give rise to which elements of the \textit{naive} activation matrix. We use the term ``naive'' since we   have not imposed reparameterization invariance at this stage; doing so is a rather complicated process which we consign to its own section, but the net effect is to remove some components of the naive activation matrix components. We    have already imposed the second order field equation conditions (\ref{eq:sec:conds-sofeqs}).

Using (\ref{eq:sec:h-defn-hl-ht}) to decompose the Fourier mode $H_{\mu\nu}$ into its longitudinal   $\qsubrm{H}{L}$ and transverse $\qsubrm{H}{T}$ scalar modes, the Eulerian perturbed energy-momentum tensor (\ref{eq:sec:emt-letter-ex}) is given by
\bea
e^{-\ci kx} \ep U^{\mu\nu} &=&  \mathsf{Y}^{(0)\mu\nu}\vphi+  \mathsf{Y}^{(1)\mu\nu}\dot{ \vphi}+  \mathsf{Y}^{(2)\mu\nu}\ddot{ \vphi}\nonumber\\
&& + \tfrac{1}{3}\gamma_{\alpha\beta}\mathsf{W}^{(0)\mu\nu\alpha\beta} \qsubrm{H}{L} + \hat{k}_{\rho}\hat{k}_{\sigma}{\perp^{\rho\sigma}}_{\alpha\beta}\mathsf{W}^{(0)\mu\nu\alpha\beta} \qsubrm{H}{T}\nonumber\\
&&+  \tfrac{1}{3}\gamma_{\alpha\beta}\mathsf{W}^{(1)\mu\nu\alpha\beta} \qsubrm{\dot{H}}{L}+\hat{k}_{\rho}\hat{k}_{\sigma}{\perp^{\rho\sigma}}_{\alpha\beta}\mathsf{W}^{(1)\mu\nu\alpha\beta} \qsubrm{\dot{H}}{T}  \nonumber\\
&&+ e^{-\ci kx}\Delta_{\{\xi\}}\ep U^{\mu\nu}.
\eea
Using the projectors (\ref{eq:sec:projectors}), the naive set of  fluid variables will be given by
\bse
\label{:eq:sec:naive-propt}
\bea
\delta \rho &=& \kappa_{11}\vphi + \kappa_{12}\dot{\vphi} + \kappa_{14}\qsubrm{H}{L} + \kappa_{15}\qsubrm{\dot{H}}{L} + \kappa_{16}\qsubrm{H}{T} + \kappa_{17}\qsubrm{\dot{H}}{T},\\
k(\rho+P)\tis &=& \kappa_{21}\vphi + \kappa_{22}\dot{\vphi} + \kappa_{24}\qsubrm{H}{L} + \kappa_{25}\qsubrm{\dot{H}}{L} + \kappa_{26}\qsubrm{H}{T} + \kappa_{27}\qsubrm{\dot{H}}{T},\\
3\delta P &=& \kappa_{31}\vphi + \kappa_{32}\dot{\vphi}+\kappa_{33}\ddot{\vphi}\nonumber\\
&&+ \kappa_{34}\qsubrm{H}{L} + \kappa_{35}\qsubrm{\dot{H}}{L} + \kappa_{36}\qsubrm{H}{T} + \kappa_{37}\qsubrm{\dot{H}}{T},\\
\pis &=& \kappa_{41}\vphi + \kappa_{42}\dot{\vphi}+\kappa_{43}\ddot{\vphi} \nonumber\\
&&+ \kappa_{44}\qsubrm{H}{L} + \kappa_{45}\qsubrm{\dot{H}}{L} + \kappa_{46}\qsubrm{H}{T} + \kappa_{47}\qsubrm{\dot{H}}{T},
\eea
\ese
where the time and space dependant \textit{naive} activation coefficients $\kappa_{\rm IJ}$ are given by
\bse
\bea
\kappa_{11} &=& u_{\mu}u_{\nu}  \mathsf{Y}^{(0)\mu\nu} ,\\
 \kappa_{12} &=& u_{\mu}u_{\nu} \mathsf{Y}^{(1)\mu\nu} , \\
 \label{kappa14}
  \kappa_{14} &=&  \tfrac{1}{3}u_{\mu}u_{\nu}\gamma_{\alpha\beta}\mathsf{W}^{(0)\mu\nu\alpha\beta}  ,\\
 \kappa_{15} &=& \tfrac{1}{3}u_{\mu}u_{\nu}\gamma_{\alpha\beta}\mathsf{W}^{(1)\mu\nu\alpha\beta},\\
 \label{kappa16}
  \kappa_{16} &=&  u_{\mu}u_{\nu}\hat{k}_{\rho}\hat{k}_{\sigma}{\perp^{\rho\sigma}}_{\alpha\beta}\mathsf{W}^{(0)\mu\nu\alpha\beta}, \\
\kappa_{17} &=& u_{\mu}u_{\nu}\hat{k}_{\rho}\hat{k}_{\sigma}{\perp^{\rho\sigma}}_{\alpha\beta} \mathsf{W}^{(1)\mu\nu\alpha\beta},
\eea
\bea
\kappa_{21} &=& \hat{k}_{\epsilon}{\gamma^{\epsilon}}_{\mu}u_{\nu}\mathsf{Y}^{(0)\mu\nu},\\
\kappa_{22} &=& \hat{k}_{\epsilon}{\gamma^{\epsilon}}_{\mu}u_{\nu}\mathsf{Y}^{(1)\mu\nu},\\
\kappa_{24} &=& \hat{k}_{\epsilon}{\gamma^{\epsilon}}_{\mu}u_{\nu}\gamma_{\alpha\beta}\mathsf{W}^{(0)\mu\nu\alpha\beta},\\
\kappa_{25} &=& \hat{k}_{\epsilon}{\gamma^{\epsilon}}_{\mu}u_{\nu}\gamma_{\alpha\beta}\mathsf{W}^{(1)\mu\nu\alpha\beta},\\
\kappa_{26} &=& \hat{k}_{\epsilon}{\gamma^{\epsilon}}_{\mu}u_{\nu}\hat{k}_{\rho}\hat{k}_{\sigma}{\perp^{\rho\sigma}}_{\alpha\beta} \mathsf{W}^{(0)\mu\nu\alpha\beta},\\
\kappa_{27} &=& \hat{k}_{\epsilon}{\gamma^{\epsilon}}_{\mu}u_{\nu}\hat{k}_{\rho}\hat{k}_{\sigma}{\perp^{\rho\sigma}}_{\alpha\beta} \mathsf{W}^{(1)\mu\nu\alpha\beta},
\eea
\bea
\kappa_{31}&=& \gamma_{\mu\nu} \mathsf{Y}^{(0)\mu\nu},\\
 \kappa_{32}&=&  \gamma_{\mu\nu}\mathsf{Y}^{(1)\mu\nu},\\
  \kappa_{33} &=& \gamma_{\mu\nu}\mathsf{Y}^{(2)\mu\nu},\\
 \kappa_{34}&=&P+ \tfrac{1}{3} \gamma_{\mu\nu}\gamma_{\alpha\beta}\mathsf{W}^{(0)\mu\nu\alpha\beta},\\
 \label{eq:sec:naive-kappa-35}
\kappa_{35} &=& \tfrac{1}{3} \gamma_{\mu\nu}\gamma_{\alpha\beta}\mathsf{W}^{(1)\mu\nu\alpha\beta},\\
  \kappa_{36} &=& P+ \gamma_{\mu\nu}\hat{k}_{\rho}\hat{k}_{\sigma}{\perp^{\rho\sigma}}_{\alpha\beta}\mathsf{W}^{(0)\mu\nu\alpha\beta},\\
 \kappa_{37} &=&  \gamma_{\mu\nu}\hat{k}_{\rho}\hat{k}_{\sigma}{\perp^{\rho\sigma}}_{\alpha\beta}\mathsf{W}^{(1)\mu\nu\alpha\beta}.
\eea
\ese
Note that $\kappa_{13}$ and $\kappa_{23}$ are not present (these would made the equations of motion higher-order). There is a clear reason we have called the $\qsubrm{\kappa}{IJ}$ the naive fluid variables: since we have not yet imposed reparameterization invariance, there should be contributions due to the components of the  $\xi^{\mu}$-field appearing in (\ref{:eq:sec:naive-propt}). 

We deliberatly have not written out the $\kappa_{4i}$: they will vanish in the reparameterization invariant theories we consider in this paper. 
The factors of $P$ appearing in $\kappa_{34}$  and $\kappa_{36}$   are due to (\ref{eq:sec:mixed-emt-withP}). We have   used (\ref{eq:sec:conds-sofeqs}) to ensure that the field equations are at most of second order. The scale dependence of the $\qsubrm{\kappa}{IJ}$ can be explicitly isolated by inspecting (\ref{eq:sectimdcoeffs-wy}) to read off the scale dependence of the $  \mathsf{Y}^{(\rm X)\mu\nu}$ and $\mathsf{W}^{(\rm X)\mu\nu\alpha\beta}$ tensors.

\subsection{Reparamerization invariance: decoupling conditions}
In order to impose reparameterization invariance, we require that the gauge field does not enter into the perturbed fluid variables. This requires 
\bse
\bea
&u_{\mu}u_{\nu} \Delta_{\{\xi\}} \ep U^{\mu\nu} = 0,\qquad u_{\mu}{\gamma^{\alpha}}_{\nu}\Delta_{\{\xi\}} \ep U^{\mu\nu}=0,&\\
& \gamma_{\mu\nu}\Delta_{\{\xi\}} \ep U^{\mu\nu}=0,\qquad {\perp^{\alpha\beta}}_{\mu\nu}\Delta_{\{\xi\}} \ep U^{\mu\nu}=0.&
\eea
\ese
These must hold at each order in $k_{\mu}$, and so
\bse
\label{eq:sec:reparam-xitheta}
\bea
&u_{\mu}u_{\nu}\Theta^{(\rm X,Y)\mu\nu}=0,\qquad u_{\mu}{\gamma^{\alpha}}_{\nu}\Theta^{(\rm X,Y)\mu\nu}=0,&\\
& \gamma_{\mu\nu}\Theta^{(\rm X,Y)\mu\nu}=0,\qquad {\perp^{\alpha\beta}}_{\mu\nu}\Theta^{(\rm X,Y)\mu\nu}=0,&\\
& u_{\mu}u_{\nu}\Xi^{(\rm X,Y)\mu\nu\sigma}=0,\qquad u_{\mu}{\gamma^{\alpha}}_{\nu}\Xi^{(\rm X,Y)\mu\nu\sigma}=0,&\\
& \gamma_{\mu\nu}\Xi^{(\rm X,Y)\mu\nu\sigma}=0,\qquad {\perp^{\alpha\beta}}_{\mu\nu}\Xi^{(\rm X,Y)\mu\nu\sigma}=0,&
\eea
\ese
where the $\Theta^{(\rm X,Y)\mu\nu}$ and $\Xi^{(\rm X,Y)\mu\nu\sigma}$ are defined in (\ref{eq:Sectcds--b43}). We call (\ref{eq:sec:reparam-xitheta}) the decoupling conditions; requiring that they hold is equivalent to requiring the theory to be reparameterization invariance.  Enforcing (\ref{eq:sec:reparam-xitheta}) upon the naive fluid variables leads to various  simplifications, which we now derive in detail. 

\subsubsection{Scale dependencies} We will pick   a few of (\ref{eq:Sectcds--b43}) to study in detail, each of which are significant. To begin with, we look at the set of coefficients,
\bea
\label{eq:sec:fdlhfsdjhfweohfewofheu}
\Theta^{(0,2)\rho\sigma} ,\qquad
\Theta^{(0,3)\rho\sigma} ,\qquad
\Theta^{(1,2)\rho\sigma} ,\qquad
\Theta^{(2,1)\rho\sigma} ,\qquad
\Xi^{(0,2)\rho\sigma\alpha}.
\eea
Applying (\ref{eq:sec:reparam-xitheta}) to (\ref{eq:sec:fdlhfsdjhfweohfewofheu}) reveals that the following tensors have no non-zero components:
\bea
\mathsf{Y}^{(0,2)\rho\sigma},\qquad
\mathsf{Y}^{(0,3)\rho\sigma}   ,\qquad
\mathsf{Y}^{(1,2) \rho\sigma}   ,\qquad
 \mathsf{Y}^{(2,1) \rho\sigma}   ,\qquad
{\gamma^{\epsilon}}_{(\mu} {\gamma^{\alpha}}_{\nu)} \mathsf{W}^{(0,1)\rho\sigma\mu\nu}.
\eea
These tensors are the coefficients of $k^2\vphi$,  $k^3\vphi$,  $k^2\dot{\vphi}$,  $k\ddot{\vphi}$ and   $kh$, $k\eta$, respectively, in all fluid variables (we   reiterate that there was never any coefficients of $k^2\ddot{\vphi}, k\dot{h}, k\dot{\eta}$); and so all terms of this form vanish from all fluid variables.

Now consider $\Theta^{(0,1)\rho\sigma}$ and apply (\ref{eq:sec:reparam-xitheta}),
\bse
\bea
\label{eq:sec:b50a}
u_{\rho}u_{\sigma}\Theta^{(0,1)\rho\sigma} & = & - u_{\rho}u_{\sigma}\mathsf{Y}^{(0,1)\rho\sigma}\dot{\phi}  - u_{\rho}u_{\sigma}\mathsf{Y}^{(1,1)\rho\sigma}\ddot{\phi}  =0    ,\\
\label{eq:sec:b50b}
\hat{k}_{\pi}u_{\rho}{\gamma^{\pi}}_{\sigma}\Theta^{(0,1)\rho\sigma} & = & - \hat{k}_{\pi}u_{\rho}{\gamma^{\pi}}_{\sigma}\mathsf{Y}^{(0,1)\rho\sigma}\dot{\phi}  -\hat{k}_{\pi}u_{\rho}{\gamma^{\pi}}_{\sigma} \mathsf{Y}^{(1,1)\rho\sigma}\ddot{\phi}  +  (\rho+P)\hat{k}_{\epsilon}{\gamma^{\epsilon\pi}} =0 ,\nonumber\\ \\
\label{eq:sec:b50c}
\gamma_{\rho\sigma}\Theta^{(0,1)\rho\sigma} & = & -\gamma_{\rho\sigma} \mathsf{Y}^{(0,1)\rho\sigma}\dot{\phi}  - \gamma_{\rho\sigma}\mathsf{Y}^{(1,1)\rho\sigma}\ddot{\phi}    =0.
\eea
\ese
(\ref{eq:sec:b50a}) and (\ref{eq:sec:b50c}) removes both $k\vphi$ and $k\dot{\vphi}$ from $\delta\rho$ and $\delta P$. If we write $\kappa_{21} = \sum \kappa_{21(n)}k^n$ and  $\kappa_{22} = \sum \kappa_{22(n)}k^n$, then  (\ref{eq:sec:b50b}) tells us that $\kappa_{21(1)}$ and $\kappa_{22(1)}$ satisfy
\bea
\kappa_{21(1)}\dot{\phi} + \kappa_{22(1)}\ddot{\phi} = \rho+P.
\eea
 
\subsubsection{Occurences of $h$ and $\eta$}
Inspecting (\ref{eq:sec:xi01}) and applying (\ref{eq:sec:reparam-xitheta}) reveals that
\bse
\bea
u_{\rho}u_{\sigma}\Xi^{(0,1)\rho\sigma\alpha}&= & 2u_{\rho}u_{\sigma}\hat{k}_{\epsilon}(\tfrac{1}{3}K{\gamma^{\alpha}}_{(\mu}u_{\nu)}    \mathbb{F}^{\epsilon\rho\sigma\mu\nu} +   {\gamma^{\epsilon}}_{(\mu}{\gamma^{\alpha}}_{\nu)}\mathbb{E}^{\rho\sigma\mu\nu}   ),\\
u_{\rho}{\gamma^{\pi}}_{\sigma}\Xi^{(0,1)\rho\sigma\alpha}&= & 2u_{\rho}{\gamma^{\pi}}_{\sigma}\hat{k}_{\epsilon}(\tfrac{1}{3}K{\gamma^{\alpha}}_{(\mu}u_{\nu)}    \mathbb{F}^{\epsilon\rho\sigma\mu\nu} +   {\gamma^{\epsilon}}_{(\mu}{\gamma^{\alpha}}_{\nu)}\mathbb{E}^{\rho\sigma\mu\nu}   ),\\
\gamma_{\rho\sigma}\Xi^{(0,1)\rho\sigma\alpha}&= & 2\gamma_{\rho\sigma}\hat{k}_{\epsilon}(\tfrac{1}{3}K{\gamma^{\alpha}}_{(\mu}u_{\nu)}    \mathbb{F}^{\epsilon\rho\sigma\mu\nu} +   {\gamma^{\epsilon}}_{(\mu}{\gamma^{\alpha}}_{\nu)}\mathbb{E}^{\rho\sigma\mu\nu}  +    P {\gamma^{\epsilon(\rho}}{\gamma^{\sigma)\alpha}}),\\
{\perp^{\zeta\pi}}_{\rho\sigma}\Xi^{(0,1)\rho\sigma\alpha}&= & 2{\perp^{\zeta\pi}}_{\rho\sigma}\hat{k}_{\epsilon}(\tfrac{1}{3}K{\gamma^{\alpha}}_{(\mu}u_{\nu)}    \mathbb{F}^{\epsilon\rho\sigma\mu\nu} +   {\gamma^{\epsilon}}_{(\mu}{\gamma^{\alpha}}_{\nu)}\mathbb{E}^{\rho\sigma\mu\nu}  +    P {\gamma^{\epsilon(\rho}}{\gamma^{\sigma)\alpha}}).\nonumber\\ 
\eea
\ese
Inside the brackets of each term is the coefficient of $h$ and $\eta$ in $\delta\rho, \theta, \delta P$ and $\Pi$ respectively, and so, by insisting that the decoupling conditions are respected, we find that $h$ and $\eta$ are not present in any fluid variables. This means that all $\qsubrm{\kappa}{IJ}$ of the form $\kappa_{i4}$ and $\kappa_{i6}$ vanish.

\subsubsection{Occurences of $\dot{h}$ and $\dot{\eta}$}
We now resolve the reparameterization-invariance conditions with respect to occurences of $\dot{h}$ and $\dot{\eta}$. To illustrate what we will be looking for, we take a simple example where the EMT given by
\bea
\delta U^{\mu\nu} = \mathbb{F}^{\rho\mu\nu\alpha\beta}\nabla_{\rho}\delta g_{\alpha\beta},
\qquad
\mathbb{F}^{\rho\mu\nu\alpha\beta} =  \mathbb{F}^{\rho(\mu\nu)(\alpha\beta)}.
\eea
We will refer to the last two indices ``$(\alpha, \beta)$'' on $ \mathbb{F}^{\rho\mu\nu\alpha\beta}$ as the ``\textit{metric indices}'' and the penultimate two indices ``$(\mu,\nu)$'' as the ``\textit{EMT-indices}''.
For this discussion it is not necessary  to know whether  these perturbations are Eulerian or Lagrangian. As usual, we   study metric perturbations in the synchronous gauge, $\delta g_{\mu\nu} = {\gamma^{\alpha}}_{\mu}{\gamma^{\beta}}_{\nu}\delta g_{\alpha\beta}$.
Using (\ref{eq:sec:nalepgdshfihsekfdsh}),  the EMT becomes
\bea
\delta U^{\mu\nu} =\big[ - u_{\rho} \mathbb{F}^{\rho\mu\nu\alpha\beta}\dot{H}_{\alpha\beta} + \ci k_{\rho} \mathbb{F}^{\rho\mu\nu\alpha\beta}H_{\alpha\beta} + \tfrac{2}{3}KH_{\rho(\alpha}u_{\beta)} \mathbb{F}^{\rho\mu\nu\alpha\beta} \big]e^{\ci kx}.
\eea
It is clear that the coefficients of $\dot{h}$ and $\dot{\eta}$ in all fluid variables are given by the spatial projection of $\mathbb{F}$ on the metric indices,
\bea
u_{\rho}{\gamma_{\alpha}}^{\pi}{\gamma_{\beta}}^{\epsilon} \mathbb{F}^{\rho\mu\nu\alpha\beta}.
\eea
The coefficients of $\dot{h}$ and $\dot{\eta}$ in the perturbed density, velocity, perturbed pressure and stress are found from application of the   projectors   defined in (\ref{eq:sec:projectors}) on the EMT-indices. We will now explicitly study the appearances of $\dot{h}$ and $\dot{\eta}$ in each of the perturbed fluid variables.

\paragraph{Appearance of $\dot{h}$ and $\dot{\eta}$ in $\tis$} First, we will prove that neither $\dot{h}$ nor $\dot{\eta}$ contribute to $\tis$. The projector of interest here is that for the scalar velocity field, $\tis$,
\bea
k(\rho+P)\tis = \hat{k}_{\epsilon}{\gamma^{\epsilon}}_{\mu}u_{\nu}\delta U^{\mu\nu}.
\eea
Hence, the coefficients of $\dot{h}$ and $\dot{\eta}$ in $\theta$ are given by the time-space projection on the EMT indices and the space-space projection on the metric indices,
\bea
u_{\rho}{\gamma^{\epsilon}}_{\mu}u_{\nu}{\gamma_{\alpha}}^{\pi}{\gamma_{\beta}}^{\zeta} \mathbb{F}^{\rho\mu\nu\alpha\beta}.
\eea
We will now prove that this vanishes, meaning that there are no occurences of $\dot{h}$ nor $\dot{\eta}$ in $\tis$. We first use $ \mathsf{W}^{(1,0)\mu\nu\alpha\beta} = \mathsf{W}^{(1,0)(\mu\nu)(\alpha\beta)}$, which is the coefficient of $\dot{h}$ and $\dot{\eta}$ in all fluid variables, and is defined in (\ref{w10-uF}). We  refer to the last two indices of $\mathsf{W}^{(1,0)\mu\nu\alpha\beta}$ as the metric-indices and the first two as the EMT-indices.
We will perform an explicit $(3+1)$ decomposition of the tensor $\mathsf{W}^{(1,0)\mu\nu\alpha\beta}$; this tensor is only a function of background quantities and so is decomposed entirely into the time-like unit-vector $u^{\mu}$ and the space-like orthogonal metric $\gamma_{\mu\nu}$ via
\bea
\mathsf{W}^{(1,0)\mu\nu\alpha\beta} =  \mathsf{W}^{(1,0)(\mu\nu)(\alpha\beta)}&=& A_{\mathsf{W}}u^{\mu}u^{\nu}u^{\alpha}u^{\beta} + B_{\mathsf{W}}u^{\mu}u^{\nu}\gamma^{\alpha\beta} + C_{\mathsf{W}}\gamma^{\mu\nu}u^{\alpha}u^{\beta} \nonumber\\
&&+ 4D_{\mathsf{W}}u^{(\mu}\gamma^{\nu)(\alpha}u^{\beta)} +E_{\mathsf{W}} \gamma^{\mu\nu}\gamma^{\alpha\beta} + 2F_{\mathsf{W}}\gamma^{\mu(\alpha}\gamma^{\beta)\nu},\nonumber\\
\eea
where the six coefficients $\{A_{\mathsf{W}}, \ldots, F_{\mathsf{W}}\}$ are background-dependant quantities. 
There are no components of $\mathsf{W}^{(1,0)\mu\nu\alpha\beta} $ which have time-space like EMT indices and space-space like metric indices. That is,
\bea
\label{eq:sec:vanish-heta-dot-theta}
u_{\mu}{\gamma_{\nu}}^{\epsilon}{\gamma_{\alpha}}^{\pi}{\gamma_{\beta}}^{\zeta}\mathsf{W}^{(1,0)\mu\nu\alpha\beta}=0.
\eea
This completes the proof that neither $\dot{h}$ nor $\dot{\eta}$ appear in $\tis$: this means that $\kappa_{25} = \kappa_{27}=0$. The key feature of the tensor $\mathsf{W}^{(1)\mu\nu\alpha\beta}$ which allowed us to do the proof in this way is that it was formed from only \textit{background} tensors -- there were no occurences of the space-like vector $\hat{k}_{\mu}$.  This observation  is also true of  (\ref{eq:sec:Y00-A}, \ref{eq:sec:Y10-fdjhfdj}, \ref{eq:sec:Y20}), so that
\bea
{\gamma^{\alpha}}_{\mu}u_{\nu}\mathsf{Y}^{(0,0)\mu\nu}=0,\qquad {\gamma^{\alpha}}_{\mu}u_{\nu}\mathsf{Y}^{(1,0)\mu\nu}=0.
\eea
These expressions would have been the coefficients of $\vphi, \dot{\vphi}$ in $\tis$ (both \textit{without} preceeding factors of $k$). Projecting these tensors (and $\mathsf{Y}^{(2,0)\mu\nu}$) with ${\perp^{\alpha\beta}}_{\mu\nu}$ also yields zero, so that $\vphi, \dot{\vphi}, \ddot{\vphi}$ do not appear in ${\Pi^{\mu}}_{\nu}$.

\paragraph{Appearance of $\dot{h}$   in $\delta P$} We now look at the occurence of $\dot{h}$ in $\delta P$, elucidated by the naive activation matrix component (\ref{eq:sec:naive-kappa-35}), and which we see is controlled by
\bea
\label{eq:sec:append-occ-doth-deltap-find}
\gamma_{\mu\nu}\gamma_{\alpha\beta}\mathsf{W}^{(1)\mu\nu\alpha\beta} =  u_{\rho}\gamma_{\mu\nu}\gamma_{\alpha\beta}\mathbb{F}^{\rho\mu\nu\alpha\beta}.
\eea
The $\mathbb{F}$-tensor is constructed from coupling tensors in the associated Lagrangian for perturbations via (\ref{eq:sec:f-defn-sol}), given by
\bea
\label{eq:sec:append-occ-doth-deltap-find-2}
\mathbb{F}^{\rho\mu\nu\alpha\beta}  = - \tfrac{1}{2} \big[ {\mathcal{U}_{}}^{\rho\alpha\beta\mu\nu}  -\mathcal{U}^{\rho\mu\nu\alpha\beta}  \big],
\eea
where we have also set $\mathcal{M}^{\epsilon\mu\nu\rho\alpha\beta}=0$ since there is no $\mathbb{G}$-term in the expansion of $\hat{\mathbb{W}}$ (\ref{eq:sec;oprators-yw-exp-defn}). Using (\ref{eq:sec:append-occ-doth-deltap-find-2}), (\ref{eq:sec:append-occ-doth-deltap-find}) becomes
\bea
u_{\rho}\gamma_{\mu\nu}\gamma_{\alpha\beta}\mathbb{F}^{\rho\mu\nu\alpha\beta}  = - \tfrac{1}{2} \gamma_{\mu\nu}\gamma_{\alpha\beta}\big[ u_{\rho}{\mathcal{U}_{}}^{\rho\alpha\beta\mu\nu}  -u_{\rho}\mathcal{U}^{\rho\mu\nu\alpha\beta}  \big]=0.
\eea
Therefore, we  conclude that $\dot{h}$ does not appear in $\delta P$, and so $\kappa_{35}=0$.

\paragraph{Appearance of $\dot{\eta}$ elsewhere} We now proceed to study the occurences of $\dot{\eta}$ in the rest of the EMT. We will start off by providing the $(3+1)$-decomposition of the tensor $\mathbb{F}^{\rho\mu\nu\alpha\beta} = \mathbb{F}^{\rho(\mu\nu)(\alpha\beta)}$,
\bea
\label{eq:sec:31demop-F-append}
\mathbb{F}^{\rho\mu\nu\alpha\beta} &=& \sbm{A}{F} u^{\rho}u^{\alpha}u^{\beta}u^{\mu}u^{\nu} + \sbm{B}{F} u^{\rho}u^{\mu}u^{\nu}\gamma^{\alpha\beta} + \sbm{C}{F} u^{\rho}u^{\alpha}u^{\beta}\gamma^{\mu\nu} + 2\sbm{D}{F}  \gamma^{\rho(\mu}u^{\nu)}u^{\alpha}u^{\beta}  \nonumber\\
&& + 2 \sbm{E}{F} \gamma^{\rho(\alpha}u^{\beta)}u^{\mu}u^{\nu}+ 4\sbm{F}{F}  u^{\rho}u^{(\alpha}\gamma^{\beta)(\mu}u^{\nu)} + \sbm{G}{F}u^{\rho}\gamma^{\alpha\beta}\gamma^{\mu\nu} + 2 \sbm{H}{F}\gamma^{\rho(\mu}u^{\nu)}\gamma^{\alpha\beta} \nonumber\\
&&+ 2\sbm{I}{F}\gamma^{\rho(\alpha}u^{\beta)}\gamma^{\mu\nu}+ 2 \sbm{J}{F}u^{\rho}\gamma^{\alpha(\mu}\gamma^{\nu)\beta}+ 4\sbm{K}{F} \gamma^{\rho(\alpha}\gamma^{\beta)(\mu}u^{\nu)} + 4 \sbm{L}{F}\gamma^{\rho(\mu}\gamma^{\nu)(\alpha}u^{\beta)} .\nonumber\\
\eea
The coefficient of $\dot{\eta}$ in all fluid variables will be found from the transverse-traceless projection on the metric indices of $\mathbb{F}^{\rho\mu\nu\alpha\beta}$, giving
\bea
\label{eq:sec:coeff-dot_eta}
{\perp^{\epsilon\pi}}_{\alpha\beta}\mathbb{F}^{\rho\mu\nu\alpha\beta} =2{\perp^{\epsilon\pi}}_{\alpha\beta} \big[  \sbm{J}{F}u^{\rho}\gamma^{\alpha(\mu}\gamma^{\nu)\beta}+ 2\sbm{K}{F} \gamma^{\rho(\alpha}\gamma^{\beta)(\mu}u^{\nu)} \big].
\eea
Notice that there will be no occurences of $\dot{\eta}$ in $\delta\rho$ (i.e. there are no time-time projections on the EMT indices, which are ``$\mu\nu$'' of the above), and $\sbm{K}{F}$ is the coefficient of $\dot{\eta}$ in $\tis$, which  we showed was zero in the proof leading up to (\ref{eq:sec:vanish-heta-dot-theta}). So,   $\dot{\eta}$ can now only appear in $\delta P$ or $\Pi$, and will only do so if $ \sbm{J}{F}\neq 0$; we will now show that reparameterization invariance enforces $ \sbm{J}{F}=0$. Earlier on, we wrote down   (\ref{eq:ec;xi11}, \ref{eq:sec:xi02}), which we repeat here, that indicate how the gauge field entered into the EMT,
\bse
\label{eq:sec:31_runup-xi11-xi02}
\bea
\Xi^{(1,1)\rho\sigma\alpha} &=&  2 ( {\gamma^{\alpha}}_{(\mu}u_{\nu)}   \mathsf{W}^{(0,1)\rho\sigma\mu\nu}+     \hat{k}_{\epsilon}{\gamma^{\epsilon}}_{(\mu}{\gamma^{\alpha}}_{\nu)}  \mathsf{W}^{(1,0)\rho\sigma\mu\nu}) ,\\
\Xi^{(0,2)\rho\sigma\alpha} &=& 2\hat{k}_{\epsilon}{\gamma^{\epsilon}}_{\mu}u_{\nu}\hat{k}_{\pi}\mathbb{F}^{\pi\rho\sigma\mu\nu},
\eea
where, repeating (\ref{w01-uF}, \ref{w10-uF}),
\bea
\mathsf{W}^{(0,1)\mu\nu\alpha\beta} \defn -\hat{k}_{\rho}\mathbb{F}^{\rho\mu\nu\alpha\beta},\qquad
\mathsf{W}^{(1,0)\mu\nu\alpha\beta} \defn u_{\rho}\mathbb{F}^{\rho\mu\nu\alpha\beta} .
\eea
\ese
All projections of $\Xi^{(1,1)\rho\sigma\alpha}$ and $\Xi^{(0,2)\rho\sigma\alpha}$ on   their first two indices must vanish for reparameterization invariance to be manifest.   We now write (\ref{eq:sec:31_runup-xi11-xi02}) using the (3+1)-decomposition (\ref{eq:sec:31demop-F-append}), yielding
\bse
\bea
\Xi^{(1,1)\mu\nu\lambda} &=& -2\hat{k}_{\rho} \big[ (\sbm{B}{F}+\sbm{E}{F} )\gamma^{\rho\lambda} u^{\mu}u^{\nu}   +( \sbm{G}{F} + \sbm{I}{F} )\gamma^{\rho\lambda} \gamma^{\mu\nu} + 2 (\sbm{J}{F} +\sbm{L}{F}) \gamma^{\rho(\mu}\gamma^{\nu)\lambda}\big],\nonumber\\ \\
\Xi^{(0,2) \mu\nu\lambda} &=& -2\big[ \sbm{E}{F}\hat{k}_{\rho} \gamma^{\rho\lambda} u^{\mu}u^{\nu}  + \sbm{I}{F}\hat{k}_{\rho}\gamma^{\rho\lambda} \gamma^{\mu\nu} + 2\sbm{L}{F}\hat{k}_{\rho}\gamma^{\rho(\mu}\gamma^{\nu)\lambda}\big] .
\eea
\ese
So, for reparameterization invariance we require, among other things,
\bea
{\perp^{\pi\zeta}}_{\mu\nu}\Xi^{(0,2) \mu\nu\lambda}  = 0,\qquad{\perp^{\pi\zeta}}_{\mu\nu}\Xi^{(1,1)\mu\nu\lambda}=0,
\eea
which implies that $\sbm{L}{F}=0, \sbm{J}{F} +\sbm{L}{F}=0$, and so clearly, $\sbm{J}{F}=0$.

This, in conjunction with (\ref{eq:sec:coeff-dot_eta}), that told us the coefficients of all occurances of $\dot{\eta}$ in the EMT, enables us to state that $\dot{\eta}$ does not appear in any components of the EMT. This means that all $\qsubrm{\kappa}{IJ}$ of the form ${\kappa}_{i7}$ vanish.

\subsubsection{The activation matrix}
After imposing (i) second order field equations and (ii) reparameterization invariance the naive activation matrix components $\qsubrm{\kappa}{IJ}$ simplify. In some instances, some of the $\qsubrm{\kappa}{IJ}$  vanish and some lose all or part of their scale dependence.  After applying all the restrictions imposed by requiring reparameterization invariance, the naive perturbed fluid variables (\ref{:eq:sec:naive-propt})  become
\bse
\bea
\delta \rho -\kappa_{15}\qsubrm{\dot{H}}{L}&=& \kappa_{11}\vphi + \kappa_{12}\dot{\vphi}    ,\\
(\rho+P)\tis &=& \kappa_{21}\vphi + \kappa_{22}\dot{\vphi}   ,\\
3\delta P &=& \kappa_{31}\vphi + \kappa_{32}\dot{\vphi}+\kappa_{33}\ddot{\vphi}    ,\\
\pis &=&0,
\eea
\ese
where all $\qsubrm{\kappa}{IJ}$ are scale independent. In the main body of the paper we write the components of the activation matrix as $\qsubrm{A}{IJ}$,   we identify
\bse
\bea
A_{11} \defn\kappa_{11},\qquad A_{12} \defn\kappa_{12},\qquad A_{14} \defn\kappa_{15},
\eea
\bea
A_{21} \defn\kappa_{21},\qquad A_{22} \defn\kappa_{22},
\eea
\bea
A_{31} \defn\tfrac{1}{3}\kappa_{31},\qquad A_{32} \defn\tfrac{1}{3}\kappa_{32},\qquad A_{33} \defn\tfrac{1}{3}\kappa_{33}.
\eea
\ese
These are precisely the same expressions as we presented in (\ref{eq:sec:fluid-vsrs-ppf}). 
\providecommand{\href}[2]{#2}\begingroup\raggedright\endgroup

\end{document}